\documentclass[lettersize,journal]{IEEEtran}
\usepackage{amsmath,amsfonts}
\usepackage{algorithmic}
\usepackage{algorithm}
\usepackage{array}
\usepackage[caption=false,font=normalsize,labelfont=sf,textfont=sf]{subfig}
\usepackage{textcomp}
\usepackage{stfloats}
\usepackage{url}
\usepackage{verbatim}
\usepackage{graphicx}
\usepackage{cite}
\usepackage{multirow}
\usepackage{array}
\usepackage{tablefootnote}
\usepackage{pifont}
\usepackage{colortbl}
\usepackage{booktabs}
\usepackage{multirow}
\usepackage{tabularx}
\usepackage{array}
\usepackage{ragged2e}
\newcolumntype{Y}{>{\RaggedRight\arraybackslash}X}
\usepackage{xcolor}
\usepackage{cleveref}
\usepackage[numbers,sort&compress]{natbib}
\usepackage{makecell}
\begin{document}

\title{Blockchain Empowered Trustworthy Agent Networks: \\Foundations, Taxonomy, and Future Directions}

\author{
	Liehuang Zhu,~\IEEEmembership{Senior Member,~IEEE,}
	Yuhang Li,
	Tianxing Wang,
	Zhihao Chen,
	Ke Li,
	Hongyi Liu,
	Yajie Wang*,
	Lei Xu,
	Peng Jiang,~\IEEEmembership{Member,~IEEE},
	Zijian Zhang,~\IEEEmembership{Senior Member,~IEEE}
	\thanks{Liehuang Zhu, Yuhang Li, Tianxing Wang, Zhihao Chen, Ke Li, Hongyi Liu, Yajie Wang, and Zijian Zhang are with the School of Cyberspace Science and Technology, Beijing Institute of Technology, Beijing 100081, China (e-mail:liehuangz@bit.edu.cn; yuhangl@bit.edu.cn; wtx\_2026@foxmail.com; chenzh511@foxmail.com; lik400865@gmail.com; liu.hongyi@bit.edu.cn; wangyajie19@bit.edu.cn; 6120180029@bit.edu.cn; pengjiang@bit.edu.cn; zhangzijian@bit.edu.cn).}

	\thanks{Corresponding author: Yajie Wang (wangyajie19@bit.edu.cn).}}

\markboth{ieee communications surveys \& tutorials}%
{Shell \MakeLowercase{\textit{et al.}}: A Sample Article Using IEEEtran.cls for IEEE Journals}


\maketitle

\begin{abstract}
AI agents are evolving from isolated task executors into networked autonomous entities that can communicate, delegate tasks, invoke tools, access external knowledge, and participate in cross-platform service and economic workflows. This evolution gives rise to open agent networks, where heterogeneous agents owned by different stakeholders interact without naturally shared infrastructures for identity, authorization, auditability, reputation, or settlement. This survey and tutorial article reviews the literature over the period 1980--2026 on the evolution from classical multi-agent systems to open agent networks, with a particular focus on LLM-based autonomous agents, agent interoperability protocols, Internet-of-Agents infrastructures, and blockchain-enabled trust mechanisms. We first review this evolution and show how the trust boundary expands from individual execution to cross-agent, cross-platform, and cross-organizational interaction. We then identify a network-level trust crisis that cannot be fully addressed by single-agent safety mechanisms or closed multi-agent coordination techniques, and develop a five-dimensional taxonomy covering entity and capability trust, authorization and delegation trust, information and provenance trust, coordination and group-robustness trust, and accountability and settlement trust. Based on this taxonomy, we examine how blockchain can provide shared identity, verifiable authorization, tamper-evident provenance, auditable collaboration, incentive alignment, and value settlement for trustworthy agent networks. We further synthesize the mapping between agent-network risks, trust requirements, and blockchain-enabled mechanisms, and clarify the role of blockchain as a shared trust layer rather than a replacement for agent security, semantic verification, privacy protection, or robust reasoning.
\end{abstract}


\section{Introduction}
\label{sec:introduction}

\IEEEPARstart{A}{I} agents are rapidly evolving from isolated task executors into networked autonomous entities that can perceive environments, reason over complex goals, invoke external tools, retrieve and update long-term memory, communicate with peer agents, delegate subtasks, and participate in service-oriented or economic workflows. Early autonomous-agent research mainly focused on decision-making and control in relatively bounded environments, whereas recent large language model (LLM)-based agents extend this paradigm by integrating language understanding, planning, memory, tool use, and interaction with external services~\cite{s3a_r1},~\cite{s3a_r5}. Tool-augmented and browser-assisted agents further demonstrate that language models can move beyond passive text generation and become active participants in digital environments~\cite{s3a_r7},~\cite{s3a_r9}.

This expansion of agent capabilities also enlarges the security and trust boundary of agent systems. In single-agent settings, major risks include prompt injection, unsafe planning and tool selection, memory poisoning, retrieval pollution, privacy leakage, and unauthorized execution~\cite{s3a_r10},~\cite{s3a_r17}. When multiple agents communicate and collaborate, additional risks arise from task delegation, shared context, debate, coordination, and intermediate-result exchange. Examples include malicious delegation, communication manipulation, hallucination propagation, covert collusion, privacy cascades, and correlated coordination failures~\cite{s3b_r11},~\cite{s3b_r15}. Agent security can therefore no longer be reduced to the safety of model outputs; it increasingly depends on the trustworthiness of execution chains, information flows, tool interfaces, communication protocols, and collaborative processes.

The next stage of agent development, however, extends beyond securing individual agents or improving cooperation within closed multi-agent systems (MAS). Agents are increasingly expected to operate in open networks, where heterogeneous agents developed by different providers, deployed on different platforms, and controlled by different stakeholders can dynamically discover one another, exchange information, invoke services, delegate tasks, and settle value across organizational boundaries. This trend is reflected in emerging interoperability protocols such as the Model Context Protocol (MCP), Agent Communication Protocol (ACP), Agent-to-Agent Protocol (A2A), and Agent Network Protocol (ANP), as well as broader visions of the Internet of Agents~\cite{S2_r3},~\cite{S2_r18}. These developments indicate a transition from platform-controlled multi-agent systems toward open, cross-platform, and economically mediated agent ecosystems.

Unlike closed multi-agent systems, participants in an open agent network cannot be assumed to share a common identity provider, authorization policy, audit mechanism, reputation system, incentive rule, or settlement infrastructure. Agents may enter and leave dynamically, advertise unverifiable capabilities, operate under different administrative domains, and delegate tasks through multi-hop interaction chains. An agent's output may subsequently become another agent's instruction, evidence, memory, tool input, or basis for payment. Consequently, a locally generated error or malicious action can propagate across agents, platforms, and time, transforming an individually containable failure into a compositional and persistent network-level problem.

We refer to this emerging problem as the trust crisis of agent networks. It arises not merely because open networks contain more agents, but because the relationships among agents, users, tools, protocols, data providers, platforms, and economic stakeholders are not supported by naturally shared trust foundations. A service requester may be unable to verify whether a discovered agent possesses its claimed identity or capability; a delegated task may be executed beyond its authorized scope; retrieved knowledge, tool outputs, or intermediate results may be manipulated without verifiable evidence; malicious or low-quality participants may strategically exploit collaboration protocols; and even a successfully completed task may lack reliable records for payment, reputation updates, dispute resolution, or responsibility attribution. These are relational risks created by networked interaction rather than independent vulnerabilities located inside a single model.

To structure this problem, this survey organizes the trust crisis of open agent networks into five interconnected dimensions. First, entity and capability trust concerns whether an agent's identity, ownership, role, and declared capabilities can be verified before interaction. Second, authorization and delegation trust concerns whether agents act only within the permissions granted by users, applications, or upstream agents and whether these permissions remain valid throughout multi-hop execution. Third, information and provenance trust concerns whether messages, retrieved knowledge, tool outputs, and intermediate evidence can be traced, verified, and audited. Fourth, coordination and collective-decision trust concerns whether collaborative processes can resist malicious, selfish, compromised, or low-quality participants. Fifth, accountability and value-settlement trust concerns whether task completion, payment, reputation updates, dispute handling, and responsibility attribution can be reliably enforced after interaction. Together, these dimensions define the network-level trust boundary that distinguishes open agent ecosystems from conventional single-agent and closed multi-agent systems.

Blockchain provides a potentially useful foundation for addressing this trust crisis, not because agent reasoning and computation should be moved entirely on-chain, but because they can offer a shared trust, incentive, and governance substrate among mutually distrustful participants. Decentralized identifiers, verifiable credentials, and distributed registries can support agent identity binding, capability declaration, and service discovery~\cite{s3c_r8},~\cite{s3c_r9}. Smart contracts can encode authorization rules, delegation constraints, task commitments, payment conditions, and dispute procedures~\cite{s3c_r17},~\cite{s3c_r19}. Tamper-evident ledgers can preserve interaction records, provenance evidence, and accountability traces, while reputation, staking, insurance, token incentives, and decentralized governance mechanisms can influence agent behavior in open service markets.

Nevertheless, blockchain should not be treated as a universal substitute for agent security. It cannot independently determine whether an LLM's reasoning is correct, whether retrieved information is semantically truthful, whether a tool invocation is safe, or whether a multi-agent decision is robust. Privacy-preserving computation, runtime monitoring, trusted execution, robust learning, semantic verification, access control, and conventional cybersecurity mechanisms remain necessary. The principal value of blockchain lies instead in transforming selected trust objects---including identities, credentials, authorization states, provenance records, task commitments, reputation updates, and settlement conditions---into verifiable states that can be recognized across organizational and platform boundaries. The important research question is therefore not simply whether blockchain can be combined with AI agents, but which network-level trust problems require decentralized verification and how blockchain mechanisms should be integrated with off-chain agent execution.

Existing surveys provide valuable foundations but remain fragmented across different system boundaries. Surveys on individual AI agents primarily examine local attack surfaces, including prompt injection, jailbreaks, unsafe planning, tool misuse, memory poisoning, retrieval pollution, privacy leakage, and unauthorized actions~\cite{S2_r1},~\cite{S2_r7},~\cite{S2_r8},~\cite{S2_r10}. Broader studies of agentic security, cyberattacks against LLM-based agents, and autonomous-agent architectures provide taxonomies of attack techniques, defense strategies, and execution pipelines~\cite{S2_r11},~\cite{S2_r12},~\cite{S2_r13}. Although these studies are essential for understanding agent-level vulnerabilities, their primary unit of analysis remains the individual agent. They do not fully capture the risks produced when heterogeneous agents interact, delegate tasks, exchange intermediate results, and operate across distinct trust domains.

Surveys on LLM-based multi-agent systems move beyond individual agents by examining agent profiling, role assignment, communication, collaboration, workflow construction, capability development, problem solving, and world simulation~\cite{S2_r2},~\cite{S2_r14}. However, these studies are generally architecture- or capability-oriented: they focus on how agents cooperate to improve task performance rather than how independently owned agents authenticate each other, delegate permissions, preserve evidence, attribute responsibility, or settle incentives. More recent studies of agent communication protocols and Internet-of-Agents infrastructures analyze MCP, ACP, A2A, ANP, agent discovery, semantic communication, task orchestration, and ecosystem governance~\cite{S2_r3},~\cite{S2_r15},~\cite{S2_r17},~\cite{S2_r18}. Protocol-level threat models and Internet-of-Agents security surveys further identify risks in authentication, access control, component provenance, cross-agent trust, supply-chain integrity, privacy, and operational reliability~\cite{S2_r4},~\cite{S2_r16},~\cite{S2_r19}. These works are closer to a networked perspective, but they mainly explain how agents interoperate or why agent networks are vulnerable; they do not systematically connect these risks to decentralized trust infrastructures.

A separate line of research surveys the intersection of AI agents and blockchain. Existing studies discuss how AI agents can assist blockchain systems through autonomous monitoring, security analysis, transaction execution, decision-making, and consensus optimization, as well as how blockchain can improve transparency and auditability in multi-agent collaboration~\cite{S2_r5}. SoK-style studies and recent surveys of autonomous on-chain agents further examine wallet interfaces, transaction intents, custody, permissioning, policy enforcement, observability, recovery, and agent-to-chain trust boundaries~\cite{S2_r6},~\cite{S2_r20}. Their emphasis, however, is commonly placed on AI-for-blockchain applications, autonomous on-chain execution, or blockchain interoperability. They do not fully generalize blockchain as a cross-organizational trust infrastructure for open agent networks spanning identity, authorization, information provenance, coordination, accountability, incentive settlement, and governance.

This article is a survey and tutorial article that surveys the literature over the period 1980--2026 on the evolution from classical multi-agent systems to open agent networks, with a particular focus on blockchain-empowered secure and trustworthy AI agent networks. Its historical scope includes classical multi-agent communication and coordination mechanisms, such as the Contract Net Protocol, KQML, FIPA specifications, and JADE~\cite{s3b_r1},~\cite{s3b_r2}. Its primary focus, however, is the recent convergence of LLM-based autonomous agents, tool-integrated agents, agent interoperability protocols, Internet-of-Agents infrastructures, and blockchain-enabled trust mechanisms. Accordingly, this survey does not attempt to cover all aspects of LLM safety, multi-agent coordination, or blockchain applications. Instead, it concentrates on the intersection where open agent networking creates cross-domain trust requirements and where blockchain can provide shared mechanisms for verification, authorization, auditability, accountability, incentive alignment, governance, and value settlement.

The main contributions of this survey are summarized as follows:

\begin{itemize}
	
	\item \textbf{A network-level security perspective and trust-crisis taxonomy.}
	We identify the risks introduced or amplified by open agent networks and organize them into five trust dimensions: entity and capability, authorization and delegation, information and provenance, coordination and collective decision-making, and accountability and value settlement.
	
	\item \textbf{A systematic analysis of blockchain-enabled trust mechanisms.}
	We examine how decentralized identity, verifiable credentials, smart contracts, audit logs, reputation and incentive mechanisms, decentralized governance, and verifiable execution can support trusted agent discovery, delegation, interaction, collaboration, accountability, and settlement.
	
	\item \textbf{An integrated risk--trust--mechanism synthesis.}
	We map agent-network risks to corresponding trust requirements and blockchain-enabled mechanisms, clarifying where decentralized verification is effective and where complementary off-chain security technologies remain necessary.
	
\end{itemize}

\section{From Single Agent to the Internet of Agents}
\label{sec:From Single Agent to the Internet of Agents}

\subsection{Single Agent Systems}
\label{subsec:Single Agent Systems}
Single-agent systems form the basic unit of agent trustworthiness. Early agents were commonly designed as autonomous entities that perceive an environment, select actions according to states, and pursue predefined goals. Their intelligence was mainly reflected in learning or executing policies within bounded environments. Reinforcement learning moved single agents from rule-based control to data-driven policy optimization, as represented by early foundational studies \cite{s3a_r1, s3a_r2}. At this stage, the environment, objective, action space, and reward function were usually well specified. Accordingly, the major trust concern was whether the agent could behave reliably within a known boundary, including risks caused by poor generalization, reward misspecification, and environmental perturbation.

The integration of large language models expanded single-agent capability from closed policy execution to open-ended task reasoning. LLMs allow agents to understand natural language instructions, interpret complex goals, and generate multi-step plans. Recent methods \cite{s3a_r3, s3a_r4} improve multi-step reasoning through explicit reasoning traces and search over alternative thoughts, while other approaches \cite{s3a_r5} connect reasoning with action and environmental observation. However, this capability also shifts the trust problem from policy stability to instruction trustworthiness and goal integrity. Because LLMs process system prompts, user inputs, and external content in the same context window, they may mistake untrusted data for executable instructions. Prior work \cite{s3a_r6} identifies this inability to separate prompt from data as a central source of prompt injection vulnerability and proposes structured queries to separate the instruction channel from the data channel.

Tool use further transforms a single agent from a language reasoning system into an external action executor. Existing research \cite{s3a_r7} shows that language models can learn when to invoke external tools; another line of work \cite{s3a_r8} demonstrates the possibility of coordinating multiple AI tools; and further frameworks \cite{s3a_r9} enable browser-assisted information access. While tools significantly improve task execution, they also move risks from textual output to real operations. Once an agent can send emails, book tickets, access banking pages, run code, or control devices, a malicious instruction may no longer merely produce an unsafe answer; it may cause unauthorized execution, privacy leakage, or real-world harm. Threat analyses \cite{s3a_r10} show that malicious instructions can be embedded in websites, emails, reviews, and other external content, inducing tool-integrated agents to harm users or exfiltrate private data. Subsequent research \cite{s3a_r11} further demonstrates that agent security must be evaluated in dynamic tool-calling environments, because tool-returned data itself can become an attack entry.

\begin{figure*}[!t]
	\centering
	\includegraphics[width=0.85\textwidth]{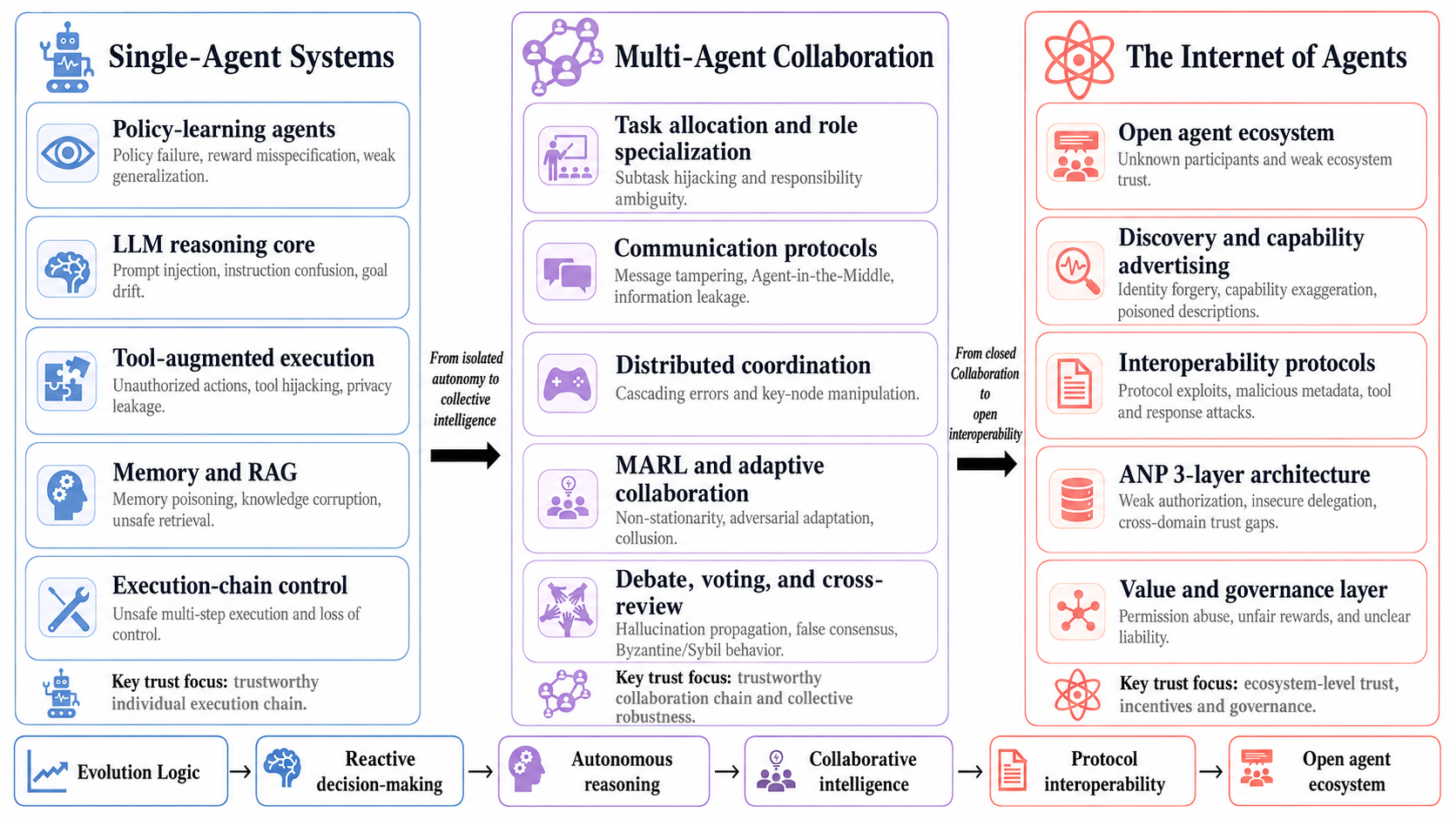}
	\caption{Overall architecture of the proposed agent network.}
	\label{fig:architecture}
\end{figure*}

When an agent is able to select tools autonomously, the trust concern further extends from safe tool execution to trustworthy tool selection. Many agent frameworks select tools according to tool descriptions, plugin documents, or retrieval results from a tool library. This makes tool documentation itself a new attack surface. Recent work \cite{s3a_r12} proposes ToolHijacker, showing that attackers can inject malicious tool documents into a tool library and manipulate the agent into selecting attacker-controlled tools for target tasks. Thus, tool augmentation is not merely a matter of capability expansion; it also creates a capability-trust problem. A tool or service may claim to provide a function, but its description, provenance, and execution result may not be trustworthy.

Memory and retrieval-augmented mechanisms then improve the long-term capability of single agents. For instance, some frameworks \cite{s3a_r13} ground generation in external knowledge to mitigate outdated knowledge and hallucination; others \cite{s3a_r14} use memory, reflection, and planning to support continuous behavior; and certain methods \cite{s3a_r15} enable self-correction through verbal feedback. Yet when agent decisions depend on long-term memory, historical demonstrations, and RAG knowledge bases, the trustworthiness of information sources becomes critical. Investigations \cite{s3a_r16} show that injecting a small number of malicious texts into a knowledge base can induce attacker-chosen answers to target questions. Further analysis \cite{s3a_r17} demonstrates that poisoning an agent's long-term memory or RAG knowledge base can influence future behavior under specific triggers. In this sense, memory is not only a module for continuity and adaptation, but also a persistent and delayed attack surface.

As planning, tool use, memory, and environmental interaction are integrated into a single execution chain, single-agent security becomes the trustworthiness of the whole individual execution process. Recent evaluations \cite{s3a_r18} use an LM-emulated tool environment to reveal long-tail failures in high-risk tool scenarios, including finance, file systems, IoT, and robotics. Other studies \cite{s3a_r19} further expand the attack surface to system prompts, user inputs, tool use, memory retrieval, backdoors, and mixed attacks, indicating that single-agent security cannot be evaluated by one prompt-injection setting alone. Correspondingly, defenses are moving from content filtering to execution constraints. Certain paradigms \cite{s3a_r20} reframe defense as task alignment, requiring every tool call to serve the user's original goal. Meanwhile, architectural solutions \cite{s3a_r21} embed trusted user intent, information-flow control, and least privilege into the execution architecture through abstract planning, concrete planning, and isolated execution.

The core vulnerability of isolated agents stems directly from their newly acquired execution autonomy. By shifting from constrained rule-following to LLM-driven reasoning and tool invocation, the system essentially hands over operational control to an unpredictable core. Consequently, the primary defensive objective at this microscopic level is containment---preventing a localized hallucination or prompt injection from escalating into a catastrophic failure. This dictates a rigid enforcement of execution guardrails, where verifiable inputs and strict authorization boundaries serve as the indispensable baseline for any reliable agentic behavior.

\subsection{Multi-Agent Collaboration}
\label{subsec:Multi-Agent Collaboration}

Multi-agent collaboration marks the transition from individual autonomy to collective intelligence. Early multi-agent systems emerged from distributed artificial intelligence, where the central goal was to allow multiple agents with different capabilities, information, or objectives to solve tasks that exceeded the capacity of a single agent. At this stage, collaboration mainly took the form of task allocation and protocol-based communication. Early literature \cite{s3b_r1} introduced a basic mechanism for distributed problem solving through task announcement, bidding, and assignment. Subsequent works \cite{s3b_r2, s3b_r3} further promoted standardized agent communication, while foundational frameworks \cite{s3b_r4} supported the engineering of agent platforms. However, protocolized communication also introduced early collaboration-trust problems: whether a message is authentic, whether a sender is reliable, and whether task allocation can be manipulated. If task announcements, bids, or communicative acts are forged, the system may suffer from incorrect division of labor or resource misallocation.

As task complexity increased, multi-agent systems evolved from simple allocation mechanisms to role-based division of labor and workflow collaboration. LLM-based MAS use natural language capabilities to assign agents different roles, such as planner, retriever, executor, reviewer, or coordinator. For instance, recent paradigms \cite{s3b_r5} have explored structured role-playing communication between agents. Others \cite{s3b_r6} have supported flexible multi-agent conversation orchestration. Moreover, specialized frameworks \cite{s3b_r7, s3b_r8} have transformed software development into a collaborative workflow among product managers, architects, programmers, and testers. Such role specialization improves complex task execution, but it also creates a trust crisis in task decomposition: if an upstream planning agent hallucinates, assigns an incorrect subtask, or is manipulated by malicious context, downstream agents may continue execution under a false premise, amplifying local errors along the workflow.

When multi-agent collaboration moves from fixed workflows to open dialogue, communication becomes the core of system capability and, at the same time, a new attack surface. Empirical studies \cite{s3b_r9} show how expert agents can interact to form collaborative capabilities, while theoretical models \cite{s3b_r10} examine how the number of agents affects collaboration. Communication allows agents to share intermediate reasoning, exchange evidence, and correct one another; however, it also allows one agent's hallucination, bias, or malicious instruction to be received and propagated by others. Threat models such as Agent-in-the-Middle attacks \cite{s3b_r11} show that an adversary does not need to compromise an individual agent; intercepting and manipulating inter-agent messages can be sufficient to change the system output. Thus, communication in MAS is not a neutral channel, but the infrastructure of collaborative security.

Multi-agent systems further develop debate, voting, cross-review, and consensus mechanisms to improve reasoning through multiple perspectives. Research on model debate \cite{s3b_r12} shows that it can improve factuality and reasoning through mutual critique. Yet debate is not automatically trustworthy. Critical analyses \cite{s3b_r13} indicate that when models have similar capabilities, training sources, or response patterns, debate may converge to the majority opinion, and that majority may reflect a shared misconception or common hallucination. Therefore, collective decision-making can enhance robustness, but it can also produce false consensus. The trust problem shifts from whether one agent is correct to whether the group decision process is resistant to bias, manipulation, and collusion.

The development of multi-agent collaboration also makes topology a security-relevant factor. Systems may adopt chain, tree, star, complete-graph, or dynamic-routing structures, and different topologies shape information flow, responsibility paths, and attack propagation. Network analysis in MAS \cite{s3b_r14} compares communication structures such as star, chain, tree, and graph, and evaluates not only task completion but also communication quality, planning quality, and individual contribution. However, the same topology that improves coordination may also influence attackability. Security evaluations \cite{s3b_r15} show that adversarial prompts can be optimized and distributed across communication-constrained networks, exploiting bandwidth limits, asynchronous message arrival, and topology-dependent propagation to bypass safety mechanisms. Hence, collaboration structure is not only an organizational design choice, but also a security-boundary design choice.

With expanding system scale and growing interaction rounds, error propagation and contamination emerge as distinctive risks inherent to MAS. Attacks that originate in single-agent settings, such as prompt injection, memory poisoning, and tool exploitation, may be amplified through dialogue rounds, shared context, and intermediate results. Recent modeling efforts \cite{s3b_r16} represent multi-agent interaction as an utterance graph and show how adversarial information can spread from a few compromised agents to a broader system through multi-turn communication. Its topology-guided intervention further suggests that multi-agent defense must identify not only malicious content, but also high-risk nodes and harmful information-flow edges. Compared with single-agent defense, MAS defense must answer three additional questions: who is compromised, how does the attack propagate, and which connections should be isolated.

At a higher level, multi-agent collaboration must address group robustness in the presence of unreliable or malicious nodes. Classic distributed systems literature \cite{s3b_r17} has long illustrated how faulty or adversarial participants can undermine consensus in distributed systems. Modern adaptations \cite{s3b_r18} apply this idea to natural-language-based MAS, emphasizing that practical MAS may not require global consensus, but they still need local coordination that remains robust against hallucinating or malicious agents. This shows that the goal of trustworthy collaboration is not merely to improve average performance, but to maintain acceptable collective behavior when some agents are unreliable, some messages are polluted, or some objectives are misaligned.

When these individual entities converge into collaborative networks, the security paradigm fundamentally shifts from isolated containment to the prevention of systemic contagion. The paradox of multi-agent systems lies in the fact that the very mechanisms driving collective intelligence---task delegation, peer dialogue, and consensus voting---simultaneously act as fertile vectors for Byzantine failures. An orchestrated attack or a localized error can now propagate seamlessly through communication topologies, exploiting implicit trust assumptions between nodes to trigger cascading breakdowns. Addressing this requires a departure from node-centric checks to a holistic validation of collaborative workflows, effectively insulating the shared context against malicious consensus manipulation.

\subsection{The Internet of Agents}
\label{subsec:The Internet of Agents}
Agent networks represent the open, networked, and ecosystem-level extension of multi-agent collaboration. Unlike closed MAS, an agent network does not assume that all agents are predefined by the same developer or coordinated by a fixed orchestration framework. Instead, agents from different organizations, platforms, and capability domains can dynamically join, discover one another, negotiate protocols, delegate tasks, and exchange services. This evolution inherits the earlier vision of distributed service interoperability. Foundational standards \cite{s3c_r1, s3c_r2} addressed service description, registration, and discovery in web-service systems. However, traditional services are usually deterministic interfaces whose invocation relations are manually engineered. In agent networks, agents can reason, plan, and invoke tools autonomously. Therefore, the question of ``whether agents can connect'' immediately becomes a trust question: whether the connected agent is authentic, whether its claimed capability is real, and whether the collaboration result can be verified.

As agent networks move from platform-internal collaboration to cross-platform interoperability, protocol standardization becomes a key developmental direction. For example, some initiatives \cite{s3c_r3} standardize how agents access tools and contextual resources; others \cite{s3c_r4} support structured messaging and asynchronous interaction; capability-based methods \cite{s3c_r5} enable task delegation among remote agents through capability descriptions; and comprehensive protocols \cite{s3c_r6} further target identity authentication, capability description, protocol negotiation, and agent discovery in open networks. These protocols reduce the cost of heterogeneous collaboration, but they also move the attack surface from individual agents to the protocol layer. Once tools, contexts, server responses, and capability metadata are exposed through standardized interfaces, adversaries can manipulate tool names, descriptions, parameters, and outputs. Security analyses \cite{s3c_r7} show that tool signatures, parameter requests, tool responses, and retrieval processes in MCP can all serve as attack vectors. Thus, interoperability protocols are both capability infrastructure and protocol-level trust boundaries.

When agents can be dynamically discovered and invoked, identity trust becomes the first prerequisite of agent networking. In closed MAS, participants are usually known in advance; in open networks, unknown agents may join at any time and claim certain skills. Decentralized identity provides a foundation for this setting. Self-Sovereign Identity (SSI) frameworks \cite{s3c_r8} allow entities to own resolvable and verifiable identifiers without relying on centralized identity providers, verifiable credentials \cite{s3c_r9} enable cryptographically verifiable claims about attributes or qualifications, and Web-based implementations \cite{s3c_r10} lower deployment barriers by using existing Web infrastructure. Yet identity trust does not automatically imply capability trust. However, trust researchers \cite{s3c_r11} emphasize that trust in SSI depends on assumptions among issuers, holders, and verifiers, rather than on DID strings alone. Therefore, even if an agent has a verifiable identity, it may still exaggerate capabilities, fabricate performance history, or manipulate reputation.

To make agents machine-understandable and selectable, capability description and discovery mechanisms become necessary. In ANP, the Agent Description Protocol describes an agent's basic information, capabilities, interfaces, authorization requirements, and service endpoints in JSON-LD. Semantic web technologies \cite{s3c_r12, s3c_r13} provide foundations for semantically structured and machine-readable descriptions. Active and passive discovery mechanisms then allow agents to be indexed and retrieved similarly to web pages. This improves task matching efficiency, but it also introduces a capability-registration trust crisis: malicious agents may forge service descriptions, poison directories, exaggerate capabilities, or register confusingly similar names. Therefore, capability discovery in agent networks cannot rely only on self-declared descriptions. It requires capability credentials, historical behavior records, remote attestation, or auditable execution evidence.

Beyond basic mutual discovery, agents' additional practices of task delegation and service invocation further complicate authorization boundaries. Traditional Internet authorization mechanisms such as OAuth \cite{s3c_r14} mainly allow users to authorize applications to access resources, while transport layer security \cite{s3c_r15} protects communication confidentiality and integrity. In agent networks, however, authorization is more dynamic and chained: a user may authorize a personal agent, the personal agent may delegate to a specialist agent, and the specialist agent may call third-party tools or services. Authorization is no longer a one-time access grant, but a transferable and multi-hop execution relationship. ANP's distinction between human authorization and agent authorization reflects this need: high-risk operations should not be fully signed by autonomous agents alone. Otherwise, an agent manipulated by malicious context may execute unauthorized actions under apparently legitimate authority.

Agent networks also extend multi-agent collaboration into open environments involving different domains, organizations, and interests. Tasks can be automatically decomposed, matched, outsourced, and recomposed, while collaboration relations change with task states. Such flexibility requires mechanisms for consensus, conflict resolution, and resource governance. Byzantine fault tolerance literature \cite{s3c_r16} provides a classical foundation for fault-tolerant consensus in environments with unreliable nodes, while rule-based governance studies \cite{s3c_r17} show how executable rules can be embedded into system operation. In agent networks, these ideas can support task-allocation confirmation, collaboration records, and dispute handling. At the same time, open collaboration raises risks of collusion, false contribution claims, low-quality services, and resource abuse. Thus, collaboration trust is not only about task completion, but also about whether contribution is genuine, the process is verifiable, and failures can be attributed.

With agents gaining economic agency, the agent network undergoes further evolution, shifting from a collaboration network to a value network. Agents may purchase data, invoke paid tools, rent capabilities, receive compensation, or participate in market bidding. Human-centric payment, contract, and liability infrastructures are poorly suited to high-frequency, automated, machine-to-machine microtransactions. Cryptocurrency and decentralized ledger technologies \cite{s3c_r18, s3c_r19} provide foundations for permissionless participation, programmable settlement, and decentralized ledgers. Integration studies \cite{s3c_r20} show that blockchain and smart contracts can record agent behavior, enforce rewards and penalties, and reduce deception and collusion. Recent frameworks \cite{s3c_r21} further apply on-chain registration, task allocation, reputation updates, and incentive mechanisms to decentralized LLM-MAS. Therefore, value-settlement trust becomes a distinctive issue of agent networks beyond ordinary MAS.

Finally, agent networks introduce a governance and liability trust crisis. In single-agent systems, responsibility can often be traced to a user, developer, or deployer. In closed MAS, responsibility can be analyzed within system boundaries. In open agent networks, however, an outcome may be jointly produced by multiple agents, tools, protocols, data sources, and platforms. Legal and ethical analyses \cite{s3c_r22} indicate that as delegation chains become longer, liability may emerge and diffuse through collaborative processes. Agent networks therefore require more than ordinary logs: they need verifiable, tamper-resistant, and replayable evidence chains, including digital signatures, timestamps, on-chain records, reputation histories, and dispute-resolution mechanisms.

Projecting these dynamics onto an open-ended Internet of Agents dismantles traditional security perimeters entirely. In a landscape defined by ad-hoc node discovery, cross-platform delegation, and machine-to-machine value exchange, the absence of a centralized orchestrator exposes a critical governance void. Trust can no longer be presumed by proximity; it must be cryptographically proven at every interaction layer. The sheer complexity of securing identity attestation, protocol interoperability, and tamper-proof settlement in such a trustless environment highlights the glaring inadequacy of conventional security frameworks. It is this absolute necessity for a deterministic, verifiable governance infrastructure that strictly mandates the integration of distributed ledger technologies into the future agentic web.

Synthesizing this evolutionary trajectory, a foundational paradox emerges: the architectural expansion of autonomous capabilities inherently outpaces the scalability of conventional security paradigms. As the threat locus systematically migrates from local execution anomalies to topology-wide contagion, and ultimately to the governance void of open ecosystems, traditional perimeter-based defenses are rendered fundamentally obsolete. We posit that merely patching node-level vulnerabilities is no longer a viable academic or engineering pursuit for the Internet of Agents. Instead, securing this macroscopic landscape necessitates a paradigm shift toward ``trust-by-design.'' It is this absolute void in decentralized accountability, cryptographic capability attestation, and deterministic execution that unequivocally necessitates the convergence of agentic networks with distributed ledger infrastructures.

\begin{figure*}[!t]
	\centering
	\includegraphics[width=0.88\textwidth]{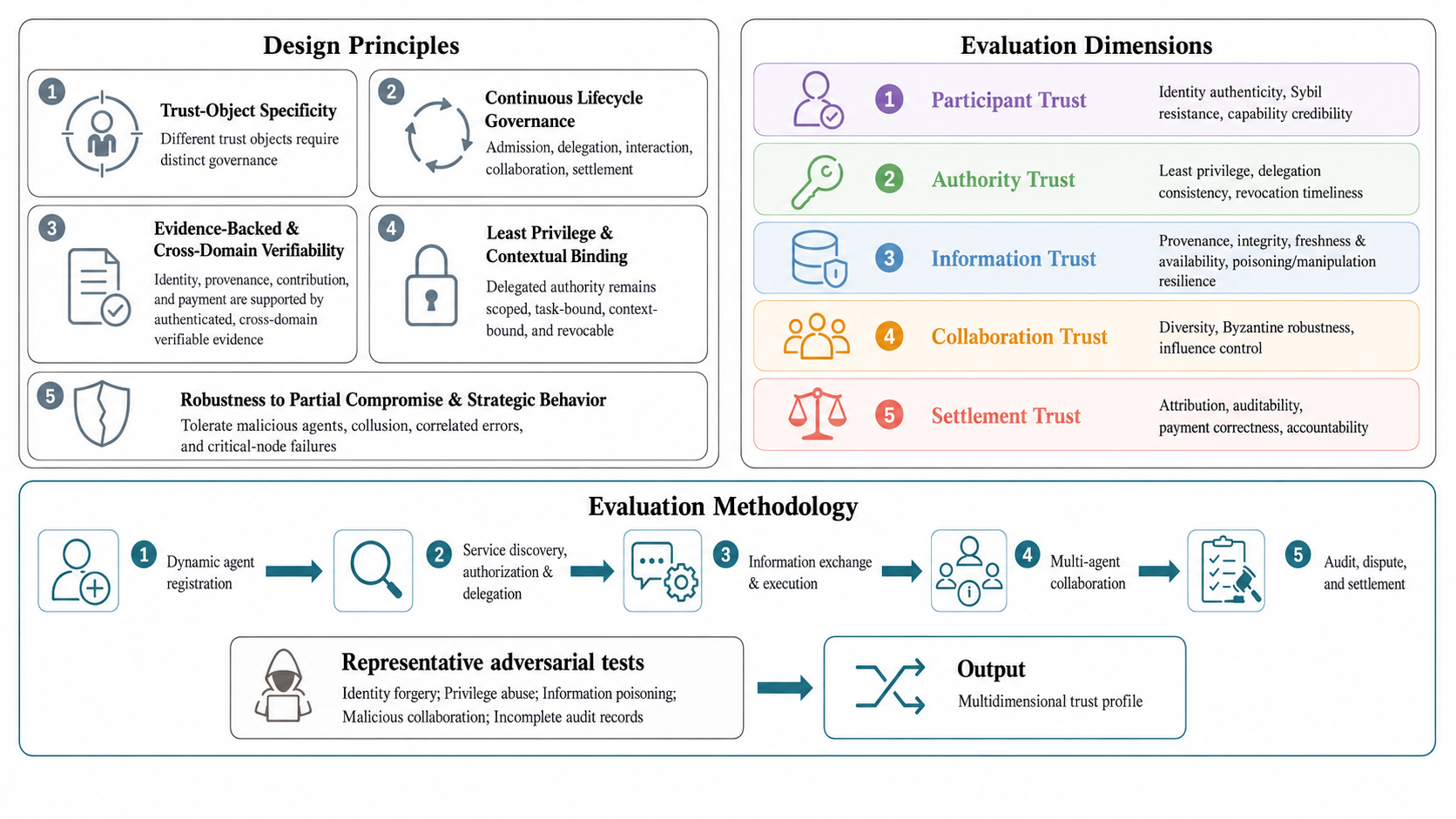}
	\caption{Design principles and evaluation framework for trustworthy open agent networks.}
	\label{fig:Crises1}
\end{figure*}

\section{Trust Crises in Agent Networks}
\label{sec:trust-crises-agent-networks}

The security risks of LLM-based agents have so far been examined mainly through the lens of prompt injection, unsafe tool use, memory poisoning, privacy leakage, and coordination failure within a single agent or a closed multi-agent system. An agent network, however, is an open, dynamic, cross-domain, and autonomous ecosystem: agents may be created by different organizations, dynamically discovered, delegated to, compensated, evaluated, and removed, with no pre-established trust relationship among participants. This openness does not merely add more agents to a fixed system; it changes the structure of risk itself. One agent's output may become another agent's input, instruction, evidence, memory, or basis for payment, turning locally containable failures into compositional, cross-agent, and temporally persistent trust crises. Yu et al.~\cite{DBLP:conf/kdd/YuMZ0MP00L00W25} show that agent trustworthiness is a system-level property spanning brain, memory, tool, user, agent, and environment modules rather than a model-only property; Tang et al.~\cite{DBLP:journals/inffus/TangLLYG26} map the attack surface of individual agents and agentic systems; and Ferrag et al.~\cite{DBLP:journals/ict-express/FerragTHMLD26} extend this surface from prompt-level threats to protocol-level risks in host-to-tool and agent-to-agent communication. Taken together, these studies indicate that no security mechanism operating solely inside one agent can fully address risks constituted by the relations among agents, tools, information sources, platforms, and stakeholders.

This network-level perspective implies several design principles for trustworthy agent networks. First, trust should exhibit trust-object specificity: participants, authority relationships, information artifacts, collaborative processes, and post-execution records are distinct objects of trust and therefore require different verification and governance mechanisms. A valid identity, for example, cannot by itself establish that an action is authorized, that a message is reliable, or that a contribution claim is accurate. Second, trustworthy operation requires continuous lifecycle governance: admission, delegation, interaction, collaboration, and settlement should be connected through an end-to-end governance process rather than secured as isolated stages. Third, trust-critical claims should satisfy evidence-backed and cross-domain verifiability. Claims concerning identity, capability, authorization, provenance, participation, contribution, and payment should be supported by evidence that can be authenticated, linked to its operational context, and examined beyond the platform that originally generated it. Fourth, authority should follow least privilege and contextual binding: delegated permissions should remain constrained by the principal's task objective, execution context, resource budget, temporal validity, sub-delegation conditions, and revocation state. Finally, the network should remain robust under partial compromise and strategic behavior, rather than assuming that all participants are honest, competent, independent, or aligned. Trustworthy operation must therefore tolerate malicious agents, low-quality services, correlated errors, collusive behavior, and failures in critical communication or coordination nodes. These principles characterize the properties that a trustworthy architecture should preserve; they do not imply that all trust failures can be addressed through a single mechanism or represented by a single trust score.

\begin{figure*}[!t]
	\centering
	\includegraphics[width=0.8\textwidth]{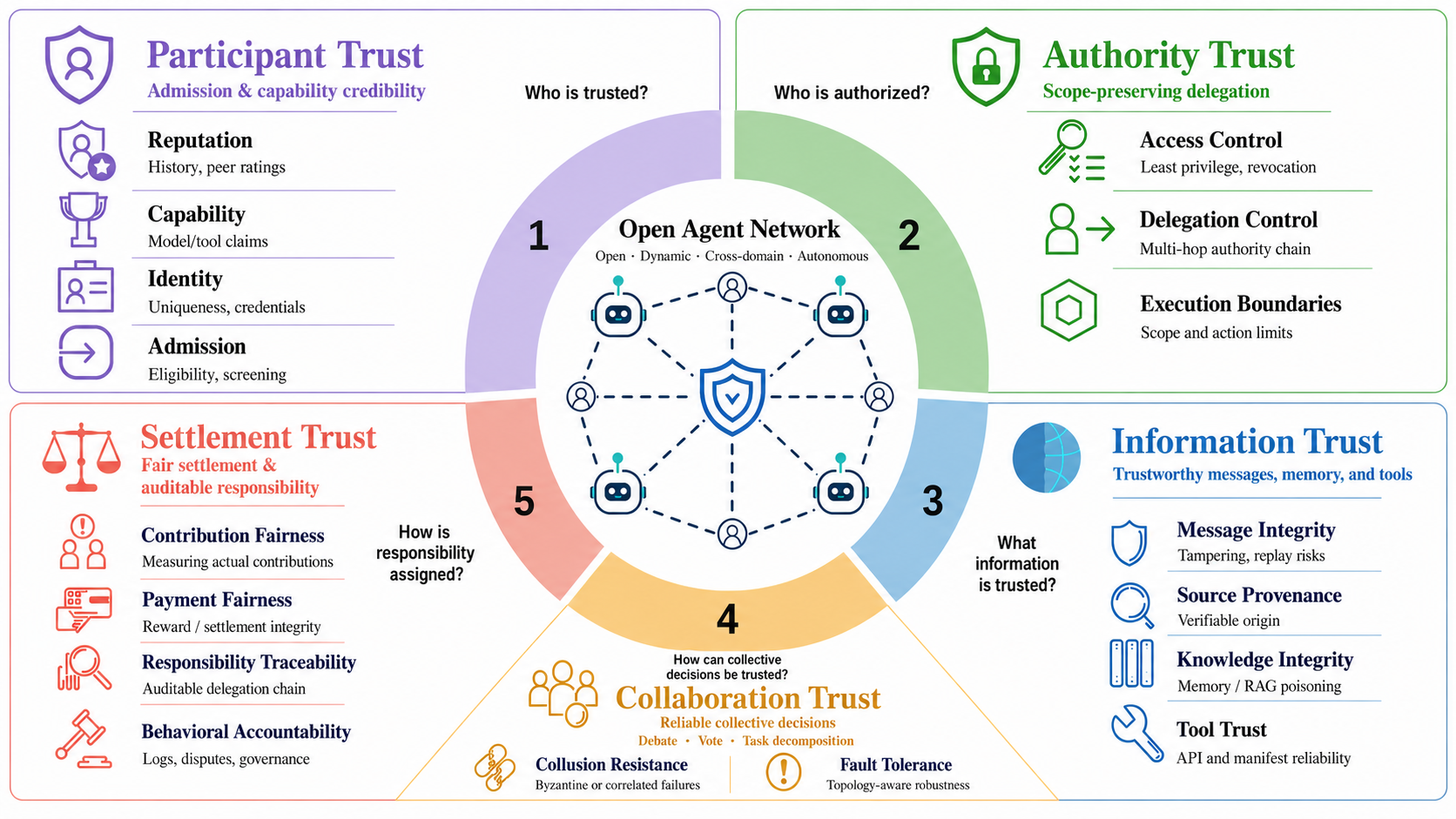}
	\caption{Trust Crises in Agent Networks.}
	\label{fig:Crises}
\end{figure*}

Guided by these principles, we organize the trust crises of agent networks along the temporal and causal arc of a delegated task, rather than around individual attack techniques. Before a task can be delegated, the network must determine whether a prospective participant should be admitted and whether its declared capability is credible---this is the crisis of entity admission and capability trust. Once admitted, the agent must operate within an authority boundary granted by a principal across planning, delegation, and tool invocation---this is the crisis of authorization, delegation, and execution-boundary trust. During execution, the agent consumes and produces messages, memories, retrieved documents, tool descriptions, environmental observations, and intermediate results whose reliability constitutes a separate object of trust---this is the crisis of information interaction and knowledge-source trust. When multiple agents jointly reason, debate, vote, allocate roles, or decompose tasks, the reliability of the resulting collective judgment becomes a further concern---this is the crisis of collaborative decision-making and group-robustness trust. Finally, after a task is completed or fails, the network must verify participation and contribution, settle value, reconstruct the execution process, and assign responsibility---this is the crisis of value settlement and accountability-governance trust.

These five crises are ordered both by the stage of the task lifecycle at which trust must be established and by the primary object of trust at that stage: the participant, the authority chain, the information artifact, the collective process, and the post-execution record, respectively. They are interconnected but analytically non-collapsible. A falsified identity is an admission failure even if no subsequent action is unauthorized; an over-broad permission is an authorization failure even if the information acted upon is accurate; a poisoned document is an information failure even if the consuming agent is legitimately admitted and properly authorized; a manipulated majority is a collaboration failure even if every individual message is authentic; and an unverifiable contribution record is a settlement failure even if the underlying task was correctly executed. The remainder of this section develops each crisis in turn, identifies its principal risk sub-types, and clarifies the boundary that separates it from adjacent crises.

\subsection{Crisis of Entity Admission and Capability Trust}
\label{subsec:entity-admission-capability-trust}

This crisis concerns whether a prospective participant should be recognized and selected by the network at all, before any task is delegated to it. Trust at this stage attaches to the participant itself --- its identity, uniqueness, declared capability, and discoverable service representation --- rather than to any action it has yet to take. The central question is whether the network has sufficient grounds to treat the participant as a legitimate and suitable node for service discovery, task routing, collaboration, or reputation formation.

Unlike a closed multi-agent system in which participants are predefined by a single developer, an open network must evaluate unfamiliar agents through machine-readable descriptions, credentials, historical signals, service manifests, capability advertisements, and possibly third-party attestations. Such evidence may be incomplete, stale, selectively disclosed, or strategically manipulated. A network may need to determine whether two visible identities belong to the same underlying controller, whether an agent's credentials are valid in another organizational domain, whether its claimed model or tool access corresponds to actual deployment, and whether its historical reputation has been earned through genuine performance rather than artificial interaction. Authorization trust becomes relevant only after an entity has already been admitted. Before that point, the more fundamental problem is whether the network should allow the entity to become a recognized participant whose claims influence discovery, selection, and future delegation.

Within this crisis, three risk sub-types can be distinguished. The first is identity and uniqueness risk: an entity may be spoofed, duplicated through Sybil behavior, or rendered unrecognizable across organizational domains. Mazzocca et al.~\cite{DIDSurvey} show that distributed identity trust depends on portable identifiers and verifiable credentials governed across domains, not on identifiers alone. This suggests that a cryptographic identifier may establish control over a key or account, but does not necessarily establish the reliability, uniqueness, institutional standing, or operational continuity of the entity behind that identifier. Crites et al.~\cite{SyRA} demonstrate that Sybil-resilient anonymous signatures are needed precisely because an adversary can otherwise generate many unlinkable identities at low cost to bias reputation, voting, or service selection. In an agent network, such behavior can distort not only voting outcomes but also routing decisions, marketplace rankings, peer recommendations, and apparent demand for a service. Krul et al.~\cite{s3c_r11} show that even verifiable identifiers depend on assumptions about issuers, holders, verifiers, and registries that may not transfer across an open network. Their analysis emphasizes that identity is not merely a technical label; it is embedded in a governance arrangement that determines which credentials are recognized, who can revoke them, and how conflicting claims are resolved.

\begin{table*}[t]
	\caption{Trust Crises in Agent Networks}
	\label{tab:trust_crises_agent_networks}
	\centering
	\footnotesize
	\setlength{\tabcolsep}{6pt}
	\renewcommand{\arraystretch}{1.15}
	\begin{tabularx}{\textwidth}{p{3.0cm}YYY}
		\toprule
		\textbf{Trust crisis} 
		& \textbf{Networked origin} 
		& \textbf{Propagation mechanism} 
		& \textbf{Network amplification} \\
		\midrule
		
		Participant Trust  
		& Open access; weak verification 
		& Untrusted agent insertion 
		& Persistent task-chain exposure \\
		
		Authority Trust 
		& Multi-hop authority 
		& Intent and permission drift 
		& Global overreach \\
		
		Information Trust 
		& Shared context and memory 
		& Reuse of polluted inputs 
		& Network-wide contamination \\
		
		Collaboration Trust 
		& Coupled reasoning 
		& Consensus amplification 
		& Correlated failure \\
		
		Settlement Trust 
		& Distributed execution 
		& Diffused responsibility 
		& Weak traceability \\
		
		\bottomrule
	\end{tabularx}
\end{table*}

The second is capability authenticity risk: a declared model backbone, tool access, domain expertise, safety profile, cost structure, latency guarantee, or task performance may not match actual behavior. LLMmap~\cite{LLMmap} shows that the underlying model of a deployed service can be fingerprinted and may diverge from what is disclosed. This creates a capability-authenticity problem because service discovery and task delegation may rely on claims about model backbone, reliability, or tool access that are not externally verifiable. An agent may advertise the use of a high-performing model while relying on a weaker substitute, may claim access to a specialized database that is intermittently unavailable, or may represent a broad tool-use capability that is only effective under narrow benchmark-like conditions.

ToolEmu~\cite{s3a_r18} and ToolLLM/ToolBench~\cite{ref9} further show that tool-use capability must be evaluated under task-specific, risk-sensitive, and API-scale conditions rather than inferred from self-description. This is particularly important in an agent network because a capability claim is not merely informational; it may determine whether the agent receives tasks with operational, financial, or privacy consequences. ShortcutsBench~\cite{ref10} and WildToolBench~\cite{ref11} likewise show that capability gaps surface under realistic workflows, compositional tasks, and implicit intent. In particular, under the benchmark conditions examined in the latter study, no evaluated model exceeded 15\% accuracy. The implication is not that all agents are necessarily unreliable, but that apparently strong tool-use performance may fail to transfer to the multi-step, underspecified, and context-dependent tasks that agent networks are expected to handle. Therefore, capability trust should be based on ongoing, task-relevant, and independently verifiable evidence rather than on static claims in a service profile.

The third is discovery-metadata manipulation: even before an agent acts, the artifacts by which it is discovered and selected can themselves be adversarial. In an open network, service descriptions, tool manifests, API documentation, capability embeddings, reputation summaries, and registry records may shape which participants become visible to other agents. ToolHijacker~\cite{s3a_r12} shows that malicious documents inserted into a tool library can steer retrieval and selection toward an attacker-chosen service without touching the consuming agent's reasoning at all. This finding is significant because it demonstrates that discovery itself is a security-critical process. A downstream agent may appear to make a rational choice based on retrieved metadata, while the retrieval environment has already been manipulated to privilege a malicious or unsuitable service. The attack therefore occurs before execution and before conventional runtime defenses can intervene.

The crisis of entity admission and capability trust is therefore the risk that the network accepts and selects an unqualified, duplicated, impersonated, or strategically misrepresented participant. Its boundary is the moment of recognition and selection: it asks whether an entity should have entered the network and been treated as a credible candidate for delegation, not what it is permitted to do once inside, nor whether the information it later exchanges is reliable. A correctly admitted agent may still violate authorization or information trust; conversely, no degree of execution-time monitoring can fully repair a network that has already admitted an illegitimate participant as a trusted node in service discovery, capability matching, and reputation formation.

\subsection{Crisis of Authorization, Delegation, and Execution-Boundary Trust}
\label{subsec:authorization-delegation-execution-boundary-trust}

The central question in this crisis is whether authority remains traceable, constrained, and semantically aligned throughout the delegation chain, from the initiating user or principal to the final operational act. Once an agent has been admitted into the network, it may receive credentials, access tokens, delegated tasks, tool permissions, data access, or the ability to issue requests to other agents. The presence of formal authorization, however, does not guarantee that every subsequent action is legitimate. A legitimate, capable, and properly admitted agent may still misuse its authority, misunderstand the scope of a task, or transfer permissions in ways that exceed the principal's actual intent.

This issue is especially important in an agent network because authority can become transitive and multi-hop. A user may instruct one agent to complete a task; that agent may delegate a component to a specialist; the specialist may invoke external tools; and those tools may access third-party services or data sources. Each local step may appear permissible in isolation, while the overall chain produces an action the original principal would not have approved. The relevant question is therefore not only whether the final actor possessed a valid token or API permission, but whether there exists an intact and legitimate authority path linking that final action back to the user's original purpose.

Likewise, the concern here is not whether a piece of content is true, but whether that content is permitted to influence a privileged action. An agent may receive accurate information from an external source, yet still lack authority to act on it. Conversely, an agent may hold legitimate permissions, but be induced by malicious or irrelevant content to use them for an illegitimate purpose. This distinction separates authorization trust from information trust: the former concerns the legitimacy of action, whereas the latter concerns the reliability of the information being acted upon.

Three risk sub-types recur in the research. The first is induced misuse of legitimately granted authority, in which an already-authorized agent is redirected by untrusted content rather than by any defect in its formal permissions. AgentDojo~\cite{s3a_r11} and InjecAgent~\cite{s3a_r10} show, across realistic tasks such as e-mail management, banking, and travel booking, that tool-returned or externally embedded content can hijack an agent that already holds useful privileges. The resulting harm does not require the attacker to steal credentials or bypass access control in the conventional sense. Instead, the attacker exploits the fact that the agent has been given authority for legitimate reasons and can be persuaded to apply that authority to an unintended objective.

ConfusedPilot~\cite{ConfusedPilot} shows the same dynamic in retrieval-augmented systems, where malicious text or cached retrieval can cause an agent to exercise authority on the basis of instructions the principal never issued. AgentFuzz~\cite{ref16} formalizes this as a taint-style vulnerability in which untrusted input flows through complex execution paths into security-sensitive operations. These studies collectively illustrate that authorization failure may emerge through the interaction of language understanding, planning, retrieval, and execution. The permissions themselves may remain technically valid, but the decision to invoke them becomes corrupted by information that should never have been allowed to control the action.

The second is the architectural conflation of instruction and data, and of planning and execution, which makes the first sub-type structurally likely rather than incidental. When an agent processes all incoming text through a common natural-language channel, it may struggle to distinguish between a user's legitimate objective, a retrieved document, a tool response, a third-party message, and a maliciously embedded instruction. ACE~\cite{s3a_r21} separates abstract planning, concrete application mapping, information-flow verification, and execution into distinct stages precisely because collapsing them into one natural-language channel is itself a source of boundary failure. StruQ~\cite{s3a_r6} addresses the same conflation by structurally separating trusted instructions from untrusted data. These approaches indicate that authorization cannot be protected solely by adding more filters at the point of action; the system must also preserve distinctions among the sources, meanings, and permissible operational roles of information throughout the workflow.

The third is semantic-task misalignment across delegation chains --- authority laundering --- in which an action is technically permitted by a coarse access token or by each individual step in a chain, yet collectively departs from the principal's original intent. A task may be decomposed into benign-looking sub-actions, each of which is authorized locally, while their combination produces a result that would not have passed a direct user-intent check. Task Shield~\cite{s3a_r20} reframes defense as verifying whether each instruction and tool call actually serves the user-specified goal, showing that permission validity and task validity are not the same property. This distinction is essential for open agent networks because a principal typically delegates an objective, not an unlimited set of possible actions that an agent might infer to be instrumentally useful.

The crisis of authorization, delegation, and execution-boundary trust is therefore the risk that an admitted participant acts without a legitimate authority path, or beyond the semantic boundary of the principal's intent, even while remaining within formally granted permissions. Its boundary is the execution chain from delegation to operational act: it does not ask whether the agent should have been admitted, nor whether the content it consumed was independently reliable, but whether a specific action was properly authorized from beginning to end. Effective governance at this layer requires more than static access control. It requires delegation-aware authority tracking, scope preservation, revocation mechanisms, and task-level verification that can identify when formally valid permissions are being applied to substantively invalid objectives.

\subsection{Crisis of Information Interaction and Knowledge-Source Trust}
\label{subsec:information-interaction-knowledge-source-trust}

Here, trust attaches primarily to the provenance, integrity, persistence, semantic reliability, and operational relevance of the information artifacts that agents exchange, store, retrieve, transform, and reuse --- messages, memories, retrieved documents, tool descriptions, environmental observations, intermediate plans, and evaluation outputs. In an agent network, information is not merely descriptive. It can directly influence planning, service discovery, tool selection, permission use, collective judgment, reputation formation, and payment. A message sent by one agent may be treated by another as evidence; a retrieved document may become a planning constraint; a tool description may determine which service is invoked; and a memory entry may shape future behavior long after its original source has been forgotten.

An agent may remain entirely within its granted permissions while still acting on corrupted, incomplete, stale, or unverifiable information. The resulting harm may be produced by an agent that is legitimate, properly authorized, and technically functioning as intended, but whose operational inputs have been manipulated. At the same time, this crisis differs from collaborative decision-making because it concerns what a single artifact contains, how it is authenticated, and how it propagates through the network, rather than how multiple agents jointly aggregate many such artifacts into a decision. It also differs from entity admission because an otherwise trustworthy sender may still transmit information that is inaccurate, compromised in transit, or misrepresented by an intermediary.

Three risk sub-types organize this layer. The first is communication-channel risk, in which the channel between legitimately admitted and authorized agents becomes an independent attack surface. The Agent-in-the-Middle attack~\cite{s3b_r11} shows that a multi-agent system can be compromised by intercepting and manipulating inter-agent messages without compromising any individual agent. This finding is especially important in open networks because the trustworthiness of each endpoint does not ensure the integrity of the path between them. A malicious intermediary may alter requests, omit warnings, inject misleading instructions, reorder messages, or selectively relay information in ways that change downstream behavior without appearing as a direct compromise of any participant.

MultiAgentBench~\cite{s3b_r14} shows that the consequences of a given corrupted message depend on communication topology --- star, chain, tree, or graph. A message injected near a central coordinator may affect many downstream agents, whereas an identical message injected at a peripheral node may remain localized. This means that information trust cannot be evaluated only at the level of message content; it must also account for the network structure through which that content travels. Kong et al.'s survey of agent communication~\cite{S2_r15}, spanning user-agent, agent-agent, and agent-environment interaction under protocols such as MCP and A2A, further shows that interoperability standards expand interaction capacity at the cost of new authentication, integrity, and endpoint-validation risks. Standardized communication can enable scalable collaboration, but it also expands the number of interfaces, schemas, connectors, and trust transitions that must be secured.

The second is persistent knowledge-base and memory poisoning, in which corrupted information outlives the interaction that introduced it. Unlike transient message manipulation, persistent poisoning changes the informational environment in which later tasks are executed. AgentPoison~\cite{s3a_r17} and PoisonedRAG~\cite{s3a_r16} show that a small number of strategically placed entries in long-term memory or a shared knowledge base can induce attacker-chosen behavior on triggered inputs while leaving benign behavior largely intact. This selective nature makes such attacks particularly difficult to detect: the affected system may appear normal under routine evaluation, while producing manipulated outcomes only under specific contextual conditions.

Yang et al.~\cite{ref25} show that such triggers need not originate in user input at all, but can be embedded in intermediate observations returned by the external environment. This expands the threat model substantially. In an agent network, environmental observations may be produced by external APIs, sensors, websites, databases, partner agents, or workflow tools. If any such source can introduce a persistent trigger into memory or retrieval infrastructure, the eventual harmful action may occur much later and be difficult to connect to the original injection event. Provenance is therefore essential not only for the authenticity of a current message, but also for understanding how a memory entry or retrieved document came to influence the system over time.

The third is protocol-level and retrieval-pipeline manipulation, in which the mechanisms of discovery and retrieval are themselves targeted. MCP Security Bench~\cite{s3c_r7} documents name collision, preference manipulation, tool-description injection, user-impersonating responses, and retrieval injection as a connected family of protocol-layer attacks. These attacks demonstrate that the information layer includes more than ordinary text content. It also includes the naming conventions, metadata fields, tool schemas, retrieval rankings, and response formats through which agents identify services and interpret external outputs. Chang et al.~\cite{ref27} show that optimized trigger fragments can make adversarial content reliably retrievable under natural queries. The RAG-jamming attack~\cite{ref28} further shows that a single blocker document can degrade answer availability even without forcing any specific false answer. Information trust, therefore, includes availability and contestability as well as factual accuracy. A system may be harmed not only by receiving false evidence, but also by being prevented from accessing relevant competing evidence.

The crisis of information interaction and knowledge-source trust is therefore the risk that polluted, unverifiable, incomplete, stale, or strategically positioned information becomes operational input for otherwise legitimate and properly authorized agents. Its boundary is the information lifecycle --- ingestion, storage, retrieval, transformation, transmission, and reuse. It does not ask whether the consuming agent was rightfully admitted or authorized, but whether what that agent consumed, and what it passes on to others, can be trusted on its own terms. Addressing this crisis requires mechanisms for provenance preservation, source authentication, integrity verification, uncertainty communication, retrieval transparency, and containment of persistent contamination across shared memory and knowledge infrastructure.

\subsection{Crisis of Collaborative Decision-Making and Group-Robustness Trust}
\label{subsec:collaborative-decision-making-group-robustness-trust}

In this crisis, the relevant unit of analysis is no longer an individual agent or message, but the collective mechanism through which judgments are formed. The issue is whether a network of agents can convert individually available information into a reliable collective judgment through debate, voting, peer review, task decomposition, role allocation, redundancy, or consensus formation. The appeal of multi-agent collaboration is that different agents may provide diverse knowledge, challenge one another's assumptions, distribute workload, and compensate for the limitations of any single model. Yet the same collaborative structure can also create new failure modes. Agents may share biases, defer to apparent authority, amplify confident but unsupported claims, conceal uncertainty, coordinate strategically, or become vulnerable to manipulation through the communication topology that connects them.

Even authentic, unaltered information can be aggregated into an unreliable conclusion if the aggregation process is flawed. Likewise, a group may reach consensus even when no participant independently possesses sufficient evidence for the conclusion. The existence of agreement is therefore not equivalent to the existence of justified trust. In open agent networks, this problem is intensified because agents may differ in model family, incentive structure, owner, tool access, communication privileges, and reliability. Some may be adversarial; some may be incompetent; some may be economically motivated to influence outcomes; and some may appear independent while in fact sharing the same underlying provider, training distribution, or strategic controller.

By contrast, questions of settlement and accountability concern the attribution of consequences after the fact, rather than the validity of a decision before or during action. Collaborative decision-making asks whether the network made a reliable judgment. Settlement asks whether the consequences of that judgment can later be fairly allocated, audited, and governed. The two are connected but analytically distinct.

Three risk sub-types are documented in the research. The first is the failure of consensus under partial or malicious participation. The Imperfect Byzantine Generals Problem~\cite{s3b_r18} argues that LLM-based agents, which communicate in natural language and produce probabilistic outputs, generally cannot satisfy the assumptions of classical global-consensus protocols, and instead require mechanisms tolerant of local or partial agreement among malicious or failed participants. The analysis highlights that the assumptions underlying traditional distributed-systems consensus --- deterministic behavior, well-defined messages, stable fault models, and clear agreement conditions --- do not transfer straightforwardly to language-based agents. A network may therefore need to decide not only whether agents agree, but also which degree of disagreement is acceptable, what evidence supports a local consensus, and how uncertainty should be represented when complete agreement is unattainable.

The second is correlated error masquerading as consensus. Estornell and Liu's theoretical analysis of multi-LLM debate~\cite{s3b_r13} shows that debate quality depends on response diversity: when participating models share training distributions, architectures, or reasoning biases, debate converges toward the majority position rather than toward the truth, so that apparent agreement is not independent confirmation. This risk is central to agent networks because multiple agents may appear diverse at the application layer while relying on highly similar foundation models, prompts, toolchains, or data sources. Redundancy without diversity can therefore create a false impression of robustness. Rather than correcting one another, agents may repeatedly reproduce the same misconception in different wording, increasing the perceived legitimacy of an incorrect conclusion.

The third is the exploitation of communication topology and group structure by an adversary that need not control the final decision-maker at all. G-Safeguard~\cite{s3b_r16} uses graph neural networks over multi-agent utterance graphs to detect anomalous influence and intervene topologically. This approach suggests that harmful influence can be identified not only from the content of individual messages, but also from the structural position of agents, the direction of information flow, and unusual patterns of attention or endorsement. Agents Under Siege~\cite{s3b_r15} optimizes adversarial prompts against latency- and bandwidth-constrained communication topologies in a permutation-invariant attack formulation, showing that attacks can be designed at the level of the network's communication structure rather than against any single prompt. In practice, an attacker may target bottleneck agents, exploit delayed communication, flood evaluators with selective evidence, or manipulate the order in which proposals are considered. The reliability of group judgment therefore depends partly on who speaks, who listens, when information arrives, and how influence is distributed.

The crisis of collaborative decision-making and group-robustness trust is therefore the risk that a network's aggregation and deliberation process yields a distorted collective outcome because of correlated error, malicious participation, strategic influence, or vulnerable communication topology, independent of whether any individual input was itself corrupted. Its boundary is the collective reasoning process: it does not evaluate the authenticity of any one message or tool call, but whether many such inputs, however individually reliable, can be combined into a trustworthy decision. Effective responses require mechanisms for diversity-aware aggregation, adversarially robust deliberation, influence monitoring, uncertainty calibration, topology-aware coordination, and procedures that prevent a superficial majority from being mistaken for epistemic reliability.

\subsection{Crisis of Value Settlement and Accountability-Governance Trust}
\label{subsec:value-settlement-accountability-governance-trust}

At this stage, trust depends on whether the consequences of networked action can be reconstructed through credible institutional records. The relevant object is the post-execution record --- contribution claims, transaction records, resource-consumption histories, audit trails, reputation updates, dispute evidence, and liability determinations --- rather than the decision or action itself. An agent network may reach a sound and well-aggregated decision, and individual agents may act within their authority boundaries, while the network still fails to determine who contributed what, who should be compensated, who consumed shared resources, or who should bear responsibility if harm occurs.

This challenge becomes increasingly important when agents participate in economic and organizational activity. Agents may purchase data, pay for computation, sell specialized services, subcontract work, consume shared resources, receive performance-based rewards, or influence reputation signals that determine future task allocation. In such settings, correctness of the final output is insufficient as a basis for trust. A useful outcome may conceal free-riding, hidden subcontracting, resource overconsumption, contribution inflation, manipulation of evaluation signals, or unfair reward allocation. Conversely, a harmful outcome may arise through a long chain of planning, retrieval, recommendation, delegation, tool invocation, and execution, making it difficult to identify the relevant responsible party.

Three risk sub-types structure this layer. The first is the absence of institutional infrastructure for agent-level economic agency, which creates incentives that a well-designed mechanism must anticipate rather than assume away. Tian's analysis of incentive-compatible multi-agent coordination~\cite{s3c_r20} shows that agents operating under reward-bearing mechanisms may optimize their own incentives rather than the delegated task unless the mechanism is designed to prevent this. This risk is not limited to overtly malicious behavior. Even agents following locally rational strategies may reduce overall system quality if rewards are poorly aligned with the principal's objective. For example, agents may prioritize visible short-term actions over less observable but necessary work, exploit loopholes in performance measurement, or strategically defer costly tasks to others.

Xu's discussion of the agent economy~\cite{TheAgentEconomy} identifies a structural gap between the economic roles agents increasingly perform --- purchasing data, paying for compute, negotiating services --- and their current lack of independent legal identity or payment capability. This gap complicates the governance of transactions, ownership, contractual commitments, and liability. If agents are economically active but lack a clear institutional status, it may be unclear who owns assets acquired through agent action, who is authorized to commit funds, and who is responsible for disputes arising from automated exchange. Qi et al.~\cite{s3c_r21} show that, absent transparent registration and verifiable task-allocation records, agents may simply deny participation or dispute contribution after the fact. In an open network, such denial can undermine not only payment but also future trust, since reputation and service discovery may depend on records that participants can later challenge.

The second is contribution attribution under ambiguous collective behavior: when a result is produced jointly, no single contribution is directly observable. This problem is especially difficult when agents act sequentially, share intermediate outputs, revise one another's plans, or make contributions through natural-language reasoning that is difficult to quantify. Learning with Reputation Reward~\cite{ref36} and Reputation-Filtered Reward Reshaping~\cite{ReputationFiltered} show that cooperation and fair reward allocation in decentralized settings depend on bottom-up reputation mechanisms that do not assume all peers are equally honest or competent. These studies imply that value settlement must account not only for the final outcome but also for the reliability and quality of the pathways through which that outcome was produced. Nagpal et al.~\cite{LLMCreditAssignment} show that even a centralized reward critic faces real difficulty decomposing a collective outcome into individual, natural-language-mediated contributions. The challenge is therefore not merely computational. It is also conceptual: a network must define what counts as contribution when agents provide planning, verification, retrieval, communication, execution, or error correction at different stages of a shared task.

The third is the reconstruction of responsibility after harm has occurred. Alqithami's lifecycle-aware accountability framework~\cite{ref39} shows that harmful behavior in a network may emerge gradually through repeated interaction rather than as a single violation, which ordinary point-in-time logging cannot capture. A harmful norm may develop as agents adapt to one another, optimize recurring incentives, or repeatedly defer responsibility across a delegation chain. Phiri's definition of auditability~\cite{CreatingAuditable} as the reliable recording, verification, and analysis of agent behavior shows that incomplete, mutable, or unstructured logs allow agents to deny actions or obscure delegation chains. Auditability therefore requires more than data retention. Records must be sufficiently complete, interpretable, tamper-resistant, and linked across agents so that a later investigator can reconstruct why an action occurred and which prior decisions influenced it.

Jan et al.'s monitored agentic architecture~\cite{ref41} shows that even well-logged individual agents may not yield a reconstructable account of a harm that emerged from their interaction. This is a crucial point for agent networks: accountability cannot be achieved by collecting isolated local logs if the harmful outcome is produced by the relation among agents. Gabison and Xian's principal-agent analysis of liability~\cite{s3c_r22} distinguishes harm traceable to individual agent design from harm that emerges only from collective behavior that no single agent intended. Their distinction is particularly relevant where responsibility is distributed among developers, deployers, users, service providers, tool owners, and autonomous agents. The network may need to determine whether harm resulted from a defective component, an unauthorized delegation, a poisoned information source, a flawed collective decision, or an emergent interaction pattern that no participant could fully control.

The crisis of value settlement and accountability-governance trust is therefore the risk that networked action cannot be followed by reliable contribution attribution, fair value distribution, verifiable resource accounting, or meaningful assignment of responsibility, even when the underlying decision and execution were sound. Its boundary is the post-execution stage: it does not ask whether an action was authorized or whether a decision was well-aggregated, but whether the consequences of that action can be reconstructed, disputed, settled, and governed on the basis of credible evidence. Addressing this crisis requires transparent task records, auditable delegation chains, contribution-aware reward mechanisms, dispute-resolution procedures, and governance arrangements capable of assigning responsibility across distributed and emergent forms of agency.

In summary, the five crises form a connected but non-collapsible chain, ordered by the stage of a delegated task at which trust must be established --- admission, authorization, information, collaboration, and settlement --- and by the object that is trusted at each stage --- the participant, the authority chain, the information artifact, the collective process, and the post-execution record. Failures at one stage can propagate to the next without being reducible to it. A fraudulent entity may obtain excessive authorization; excessive authorization may expose agents to corrupted information or enable unsafe use of it; corrupted information may distort collective decisions; and distorted decisions may produce harms that cannot be settled without reliable attribution.

The chain can also loop backward. A disputed settlement may corrupt the reputation signals on which future admission decisions depend. A failure to reconstruct responsibility may prevent the network from identifying which capability claims should be downgraded or which participants should be excluded. A manipulated reputation record may then influence future service discovery, task allocation, or delegation authority. Thus, accountability failures can become tomorrow's admission failures, and information failures can become persistent capability or reputation failures. Securing any single agent, or even any single layer, therefore cannot resolve a risk that is constituted by the relations among agents across the full lifecycle of delegated, networked action. A trustworthy agent network requires coordinated governance of who may participate, what they may do, what information they may rely on, how they collectively decide, and how their actions are ultimately valued, audited, and held to account.

Building upon the above analysis, the trustworthiness of an agent network should be evaluated as a multidimensional system property rather than a single security attribute. Since different trust objects correspond to different governance objectives, no individual metric is sufficient to characterize the overall trustworthiness of the network. Instead, evaluation should assess whether each trust object can be continuously established, maintained, and verified throughout the lifecycle of delegated task execution.
Accordingly, trustworthy agent networks can be evaluated along five dimensions corresponding to the five trust crises: entity admission and capability trust, authorization, delegation, and execution-boundary trust, information interaction and knowledge-source trust, collaborative decision-making and group-robustness trust, and value settlement and accountability-governance trust. For brevity, we refer to these dimensions as participant, authority, information, collaboration, and settlement trust, respectively. Participant trust measures whether entities entering the network possess authentic, distinguishable, and verifiable identities, together with reliable capability claims and valid admission credentials. Authority trust evaluates whether delegated permissions remain consistent with the principal’s intent, satisfy the principle of least privilege, preserve their contextual and temporal constraints across sub-delegation, and can be revoked or updated in a timely manner. Information trust evaluates whether the information artifacts and sources used or produced by agents—including messages, memories, retrieved documents, tool descriptions, environmental observations, protocol metadata, and intermediate results—possess verifiable provenance, protected integrity, adequate freshness and availability, and resilience to poisoning, suppression, or manipulation. Provenance and integrity should be evaluated separately from semantic correctness, because information may be authentic and unmodified while still being inaccurate, misleading, incomplete, or operationally irrelevant. Collaboration trust evaluates whether collective decisions remain reliable under heterogeneous capabilities, adversarial participants, correlated errors, communication uncertainty, vulnerable network topologies, or strategic behavior. Finally, settlement trust assesses whether participation, contribution, resource consumption, responsibility, and value distribution can be reconstructed after task execution through complete audit trails, non-repudiable evidence, and verifiable accountability mechanisms.
Correspondingly, evaluation methodologies should also move beyond isolated benchmark tests toward lifecycle-oriented assessments. Rather than measuring individual agents independently, experiments should reproduce realistic delegated workflows involving dynamic agent registration, service discovery, authorization, multi-agent collaboration, information exchange, execution auditing, and post-task settlement. Adversarial scenarios should be introduced throughout this lifecycle to evaluate the resilience of each trust object under identity forgery, privilege abuse, information poisoning, malicious collaboration, incomplete audit records, and other representative attacks. The resulting evaluation should therefore be reported as a multidimensional trust profile instead of a single trust score, enabling researchers to identify which trust objects remain vulnerable, how failures propagate across different stages of delegated execution, and which governance mechanisms are required to improve the trustworthiness of the overall agent network.

\begin{figure*}[!t]
	\centering
	\includegraphics[width=0.9\textwidth]{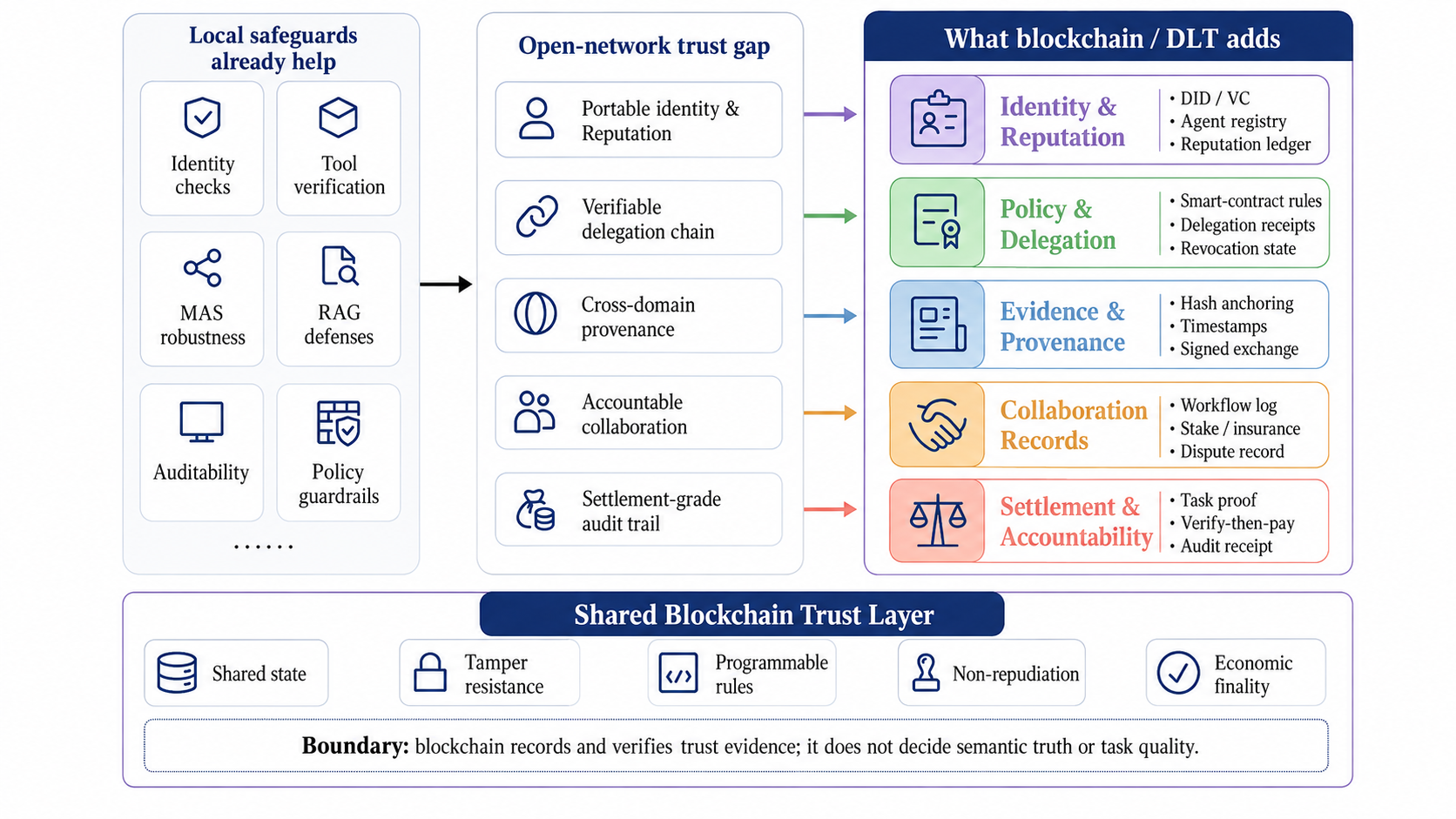}
	\caption{Blockchain as a shared trust layer and its functional boundaries in open agent networks.}
	\label{fig:Block}
\end{figure*}

\section{Blockchain-Based Trust Enhancement for Open Agent Networks}
\label{sec:part5-blockchain-based-trust-enhancement-for-open-agent-networks}

This chapter examines LLM-based agent networks under open interconnection paradigms such as A2A, IoA, the Agentic Web, and Agentic Commerce, rather than platform-internal orchestration or multi-agent collaboration within a single task boundary. The preceding section has organized risks in open agent networks into five trust crises: entity admission and capability trust, authorization, delegation, and execution-boundary trust, information interaction and knowledge-source trust, collaborative decision-making and group-robustness trust, and value settlement and accountability-governance trust. The question therefore shifts to trust infrastructure: once these crises have been identified, how can identity, authorization, information, collaboration, and settlement states be continuously verified, inherited, contested, resolved, and recorded across organizational boundaries?

Around this question, this paper positions blockchain/DLT as a supplementary trust infrastructure for open networks, focusing on its role in shared state, non-repudiable evidence, incentive constraints, and programmable settlement. Existing non-blockchain research has already established important foundations in identity credentials, prompt-injection defense, RAG robustness, and auditability \cite{DIDSurvey,s3a_r6,RobustRAG,CreatingAuditable}. These methods can reduce runtime risks in concrete systems, but most still rely on a single platform, one-time evaluation, local logs, or controlled execution environments. When agents are discovered, delegated, communicated with, coordinated, and transacted with across organizations, the trust problem is no longer only whether one system can detect a risk. It also concerns whether different parties can jointly recognize the same state, evidence, and settlement result. The supplementary value of blockchain/DLT lies here: through on-chain registration, hash anchoring, smart contracts, non-repudiable logs, staking, reputation, insurance, and micropayments, local security results can be turned into trust objects that are jointly verifiable by network participants.

\subsection{Scope and Limits of Blockchain}

In this chapter, blockchain is treated as trust infrastructure rather than a general purpose security mechanism. On-chain records can verify that identity claims, authorization states, message digests, execution proofs, or payment events existed at a given time and were not subsequently altered. However, they cannot determine whether natural language intent has been manipulated, retrieved content is truthful, tool descriptions are semantically safe, capability claims are valid, or multi-agent agreement is correct.

Within this scope, blockchain is most suitable for four types of cross-organizational problems. First, state consistency concerns the shared maintenance of identity, authorization, revocation, reputation, and payment states. Second, evidence verifiability concerns the provenance and integrity of messages, invocations, executions, and audit records. Third, behavioral constraint concerns encoding budgets, permissions, stakes, penalties, and payment conditions as executable rules. Fourth, settlement enforceability concerns connecting service completion, task proofs, compensation, and dispute handling within a common settlement process. Semantic attack defense, content assessment, capability evaluation, privacy compliance, and legal adjudication still require complementary security and governance mechanisms. Therefore, this chapter focuses on a layered trust architecture in which blockchain and non-blockchain methods serve complementary roles.

\begin{table*}[!t]
	\caption{Non-Chain Gaps, Blockchain-Based Supplements, and On-Chain Boundaries in Open Agent Networks}
	\label{tab:part5-trust-crises}
	\centering
	\renewcommand{\arraystretch}{1.15}
	\setlength{\tabcolsep}{2.5pt}
	\footnotesize
	\def\tabitem{\textbullet\ }
	\begin{tabular}{|>{\centering\arraybackslash}p{0.12\textwidth}|>{\raggedright\arraybackslash}p{0.09\textwidth}|>{\raggedright\arraybackslash}p{0.17\textwidth}|>{\raggedright\arraybackslash}p{0.20\textwidth}|>{\raggedright\arraybackslash}p{0.12\textwidth}|>{\raggedright\arraybackslash}p{0.18\textwidth}|}
		\hline
		\textbf{Trust crisis} & \textbf{Simplified trust object} & \textbf{Non-chain gap} & \textbf{Blockchain-based supplement} & \textbf{Representative methods} & \textbf{On-chain limits} \\
		\hline
		Entity admission\newline and capability trust &
		\tabitem Identity\newline \tabitem Capability\newline \tabitem Endpoint\newline \tabitem Reputation &
		\tabitem Platform-bound accounts\newline \tabitem One-shot tests\newline \tabitem Weak cross-domain revocation\newline \tabitem Non-portable reputation &
		\tabitem DID/VC anchoring\newline \tabitem AgentCard/ANS registration\newline \tabitem Endpoint tamper evidence\newline \tabitem Credential digests\newline \tabitem Revocation state\newline \tabitem Reputation ledger &
		\tabitem BlockA2A\newline \tabitem Binding Agent ID\newline \tabitem ANS\newline \tabitem AgentReputation &
		\tabitem Anchored capability evidence\newline \tabitem Off-chain quality evaluation\newline \tabitem Cold-start reputation\newline \tabitem Reputation/Sybil attacks \\
		\hline
		Authorization,\newline delegation, and\newline execution-boundary trust &
		\tabitem Delegation chain\newline \tabitem Scope\newline \tabitem Budget\newline \tabitem Revocation &
		\tabitem Local access control\newline \tabitem Coarse API tokens\newline \tabitem Weak subdelegation tracking\newline \tabitem Fragmented budget state &
		\tabitem Smart-contract access control\newline \tabitem On-chain delegation records\newline \tabitem Revocable authorization\newline \tabitem Scoped authorization\newline \tabitem Locked budgets\newline \tabitem Conditional payment &
		\tabitem BlockA2A\newline \tabitem Agent-OSI\newline \tabitem A2A+x402\newline \tabitem Trust primitives &
		\tabitem Formal rules only\newline \tabitem No intent understanding\newline \tabitem Authorized misuse\newline \tabitem Tool attacks and drift \\
		\hline
		Information interaction\newline and knowledge-source trust &
		\tabitem Messages\newline \tabitem Metadata\newline \tabitem Versions\newline \tabitem Evidence &
		\tabitem Fragmented local logs\newline \tabitem Incomplete provenance\newline \tabitem Mutable version history\newline \tabitem Weak cross-platform evidence &
		\tabitem Message hashes\newline \tabitem Signed digests\newline \tabitem Timestamps\newline \tabitem Version chains\newline \tabitem Metadata anchoring\newline \tabitem Evidence indices &
		\tabitem DMAS\newline \tabitem MOD-X\newline \tabitem BetaWeb\newline \tabitem Trust Fabric &
		\tabitem Integrity, not truth\newline \tabitem False-content attestation\newline \tabitem Off-chain storage needed\newline \tabitem Sensitive data off-chain \\
		\hline
		Collaborative decision-making\newline and group-robustness trust &
		\tabitem Participants\newline \tabitem Topology\newline \tabitem Contribution\newline \tabitem Disputes &
		\tabitem Task-local checks\newline \tabitem Weak process inheritance\newline \tabitem Unverifiable contribution paths\newline \tabitem Unclear dispute evidence &
		\tabitem Collaboration traces\newline \tabitem Participant signatures\newline \tabitem Result commitments\newline \tabitem ZK contribution proofs\newline \tabitem Staking\newline \tabitem Insurance\newline \tabitem Dispute records &
		\tabitem Agent Exchange\newline \tabitem DAO-Agent\newline \tabitem WBFT\newline \tabitem Insured Agents &
		\tabitem Consensus, not correctness\newline \tabitem Correlated errors\newline \tabitem Incentives only\newline \tabitem Governance needed \\
		\hline
		Value settlement\newline and accountability-governance trust &
		\tabitem Task proof\newline \tabitem Payment\newline \tabitem Audit\newline \tabitem Liability &
		\tabitem Platform-local records\newline \tabitem Weak payment binding\newline \tabitem Mutable audit logs\newline \tabitem Unclear liability chain &
		\tabitem Verify-then-pay\newline \tabitem x402 micropayments\newline \tabitem Escrowed payment\newline \tabitem Task proofs\newline \tabitem Staking penalties\newline \tabitem Insurance contracts\newline \tabitem Audit ledger &
		\tabitem TessPay\newline \tabitem A2A+x402\newline \tabitem CPMM\newline \tabitem Agent TCP/IP &
		\tabitem External verification\newline \tabitem Uncertain service quality\newline \tabitem Legal liability off-chain\newline \tabitem Privacy/compliance risks \\
		\hline
	\end{tabular}
\end{table*}

\subsection{Entity Admission and Capability Trust}
\label{subsec:part5-1-entity-admission-and-capability-trust}

At the entity-admission stage, the object of trust is the agent as a network participant, including its identity, uniqueness, capability claims, service endpoint, and reputation state. Admission trustworthiness in open networks is not limited to whether an agent passes authentication at a given moment. It concerns whether the relevant states can be continuously verified, inherited, and updated across organizations.

\subsubsection{Existing local Safeguards}
\label{subsubsec:part5-1-1-existing-non-blockchain-methods}

Non-chain research has already laid important foundations for the local credibility of identity and capability claims. The survey by Mazzocca et al.~\cite{DIDSurvey} on decentralized identifiers and verifiable credentials shows that identity trust in distributed environments depends on portable identifiers, verifiable credentials, and cross-domain governance. SyRA~\cite{SyRA} constrains the number of pseudonyms in a given context through cryptographic mechanisms, thereby mitigating the effect of Sybil identity proliferation on reputation, voting, and service selection. These works show that admission trustworthiness cannot rely only on platform accounts or self-declared names; it requires verifiable identifiers, credentials, and constraints on identity multiplicity.

On the capability side, non-blockchain research can reveal model substitution, exaggerated capability claims, and tool-metadata contamination. LLMmap~\cite{LLMmap} uses active queries and response-pattern analysis to identify the model version behind LLM-integrated applications, showing that the deployed model may diverge from the public claim. ToolEmu~\cite{s3a_r18} shows that tool-use capability should be evaluated under task-relevant and risk-sensitive conditions rather than inferred from self-reported metadata. The ToolHijacker attack~\cite{s3a_r12} further demonstrates that tool descriptions and capability metadata can themselves become attack surfaces for manipulating discovery and selection processes.

\subsubsection{Residual Trust Gap}
\label{subsubsec:part5-1-2-residual-trust-gap}

These methods have direct value for identity authentication, model identification, capability evaluation, and metadata inspection, but their results often remain one-time verification events, local tests, or platform-internal records. SyRA can constrain the number of identities, but it does not indicate whether an agent's historical performance has been reliable. LLMmap can identify model inconsistency, but it does not easily convert this finding into a long-term cross-platform state. ToolEmu can evaluate capability in task environments, but it usually cannot allow other organizations to jointly confirm who passed which evaluation at what time and whether that conclusion is still valid. Open agent networks require not isolated verification events, but long-term associations among identity, capability, endpoint, and reputation that are accumulative, revocable, and contestable.

\subsubsection{Blockchain Trust Augmentation}
\label{subsubsec:part5-1-3-blockchain-based-mechanism-supplement}

In the admission stage, blockchain/DLT mainly pulls identity registration, endpoint anchoring, and state recording out of platform-internal directories and turns them into objects that can be jointly verified across organizations. BlockA2A~\cite{BlockA2A} adopts DIDs for fine-grained cross-domain agent authentication and records identity claims and interaction states through a blockchain-anchored ledger, enabling different organizations to jointly verify an agent's authentication materials and historical states. Binding Agent ID~\cite{BindingAgentID} links agent discovery, identity verification, service-condition checking, and authorization-limit enforcement, making identity an entry point for later performance, payment, and accountability rather than an isolated label. A2A+x402~\cite{A2Ax402} publishes AgentCards to DLT, making service endpoints tamper-resistant, verifiable, and settleable.

Mechanistically, an on-chain registry is not another account system. It abstracts subject states out of platform databases and turns them into public indexes that multiple parties can verify. DIDs, public keys, AgentCard hashes, capability-credential digests, revocation states, and reputation updates can be recorded in on-chain or DLT registries. Complete capability descriptions, model configurations, tool documents, and privacy-sensitive materials can remain off-chain and be bound to on-chain records through hashes, signatures, or verifiable credentials \cite{BlockA2A,ANS}. This structure allows network participants to check whether a service endpoint has been altered, whether a credential has been revoked, and whether a capability evaluation still corresponds to the current version, without fully trusting the directory or logs of a single platform.

Discovery and registration mechanisms further turn identity states into operational admission bases. Agent Name Service (ANS)~\cite{ANS} provides a universal directory compatible with standards such as A2A, MCP, and ACP, enabling service registration, resolution, and capability announcement to be securely discovered in open networks. Agent-OSI~\cite{AgentOSI} incorporates stable identifiers, authentication, and authorization into a layered protocol stack, allowing admission states to connect with later delegation, settlement, and provenance. Trust Fabric~\cite{TrustFabric} provides trust-coordination mechanisms for Agentic Web interoperability through decentralized registries, semantic Agent Cards, and verifiable credentials.

Capability and reputation make this state-oriented logic especially clear. Agent Exchange~\cite{AgentExchange} links capability representation, performance tracking, task coordination, and value attribution with market mechanisms. AgentReputation~\cite{AgentReputation} emphasizes that reputation systems require verifiable evidence, privacy protection, and manipulation resistance. Research on the Agent Economy~\cite{TheAgentEconomy} places W3C DIDs, on-chain sovereign identity, and reputation capital among the foundational conditions for agents to become economic actors. On-chain mechanisms do not directly evaluate capability. They preserve capability-evaluation results, credential digests, version changes, and reputation updates so that these results can be inherited, reviewed, and contested by later organizations.

Still, on-chain registration and reputation records only make identity, endpoint, and historical states easier to verify across domains; they do not automatically prove that a capability claim is true. Cold start, evaluation strength, and reputation manipulation still require non-chain evaluation, market feedback, and governance mechanisms \cite{AgentReputation}.

\subsection{Authorization, Delegation, and Execution-Boundary Trust}
\label{subsec:part5-2-authorization-delegation-and-execution-boundary-trust}

At the authorization-execution stage, the object of trust is the authority chain from the original delegation to the final operation, including authorization scope, subdelegation conditions, budget constraints, revocation state, and task-goal consistency. Authorization in open networks is not only a question of whether permission exists. It also concerns whether permission can be jointly recognized and constrained after it moves across subjects.

\subsubsection{Existing Non-Blockchain Methods}
\label{subsubsec:part5-2-1-existing-non-blockchain-methods}

In non-chain research, authorization defenses more often target permission misuse and unauthorized execution within a single system. StruQ~\cite{s3a_r6} distinguishes trusted instructions from untrusted data through structured inputs, representing a channel-separation approach to mitigating instruction/data confusion. ACE~\cite{s3a_r21} limits unauthorized paths through abstract planning, application mapping, information-flow verification, and execution barriers, showing that authorization boundaries cannot rely only on natural-language constraints. Task Shield~\cite{s3a_r20} checks at runtime whether tool calls serve the user's goal, further demonstrating that permission validity is not the same as task consistency.

Within a single agent, platform, or execution environment, these methods have direct effects. When the system can control input channels, tool sets, execution environments, or logging, permission misuse caused by prompt injection, confused-deputy behavior, and execution drift can be substantially reduced. The induced misuse of authority revealed by AgentDojo~\cite{s3a_r11} and ConfusedPilot~\cite{ConfusedPilot} also shows that authorization failure often does not arise because a token itself is invalid, but because an already authorized agent is guided by a complex information flow to perform an operation that deviates from the original objective.

\subsubsection{Residual Trust Gap}
\label{subsubsec:part5-2-2-residual-trust-gap}

In open agent networks, authorization relationships move continuously with task decomposition, service invocation, and tool use. One agent may receive a user delegation, delegate a subtask to another agent, and the latter may then invoke external services or tools. Each local step may be formally legitimate within its own platform, but the cross-organizational chain as a whole may lack a jointly verifiable state showing whether it still conforms to the original authorization scope. Non-chain methods can reduce semantic inducement and permission misuse during a particular execution, but they struggle to answer whether authorization has been legally inherited, whether further subdelegation is allowed, whether revocation is recognized by all participants, whether budget and invocation counts remain within scope, and whether evidence can be accepted by third parties in a dispute.

\subsubsection{Blockchain-Based Mechanism Supplement}
\label{subsubsec:part5-2-3-blockchain-based-mechanism-supplement}

Once authorization moves into a cross-subject network, the role of blockchain/DLT concentrates on recording, executing, and reviewing authorization states. BlockA2A~\cite{BlockA2A} uses smart contracts for dynamic, execution-context-aware access control, forming a closed loop among cross-domain identity authentication, ledger auditability, and policy execution in A2A interoperability. Agent-OSI~\cite{AgentOSI} treats Identity, AuthN \& Authorization as an independent protocol layer, covering stable identifiers, delegation, consent, and policy enforcement. This protocol structure understands authorization as a continuous process spanning upstream intent, agent execution, downstream services, and result return.

The most important change is that authorization is no longer only a platform-internal configuration; it can be represented as a cross-subject delegation record. The authorizing subject, authorized subject, callable services, time windows, budget ceilings, invocation counts, risk levels, subdelegation scope, and revocation state can all be encoded as contract states or policy objects. Later participants then see not only whether an API call succeeded, but also whether the call belonged to a legitimate delegation chain, whether it exceeded budget or scope, whether it occurred before revocation, and whether it satisfied service and payment conditions.

When authorization enters real service invocation, on-chain mechanisms can further couple authorization boundaries with budgets, payments, and performance conditions. A2A+x402~\cite{A2Ax402} combines on-chain AgentCards with x402 micropayments, allowing service invocation to check counterpart identity, invocation authorization, and payment conditions at the same time. A systematization study on blockchain-based A2A payments~\cite{SoKA2APayments} summarizes the payment lifecycle as discovery, authorization, execution, and accounting, and identifies insufficient intent binding and misuse under valid authorization as key challenges. A systematization study on Agentic Commerce security~\cite{SoKAgenticCommerce} also treats transaction authorization as a core risk dimension in agent commerce systems.

Authorization boundaries also require more fine-grained protocol primitives for expression and composition. Hu and Rong~\cite{InterAgentTrustModel} compare primitives such as brief, claim, proof, stake, reputation, and constraint within a unified framework, showing that trust in open networks is shifting from human supervision to protocolized constraints. Building Secure A2A~\cite{BuildingSecureA2A} also shows, from a non-chain A2A security-practice perspective, that authorization boundaries require structured description, validation, and testing. On-chain policies therefore supplement the shared visibility and enforceability of authorization relationships, rather than replacing the model's understanding of task semantics.

One basic boundary remains: on-chain policies can prove whether authorization states, budget conditions, and subdelegation scopes have been followed, but they do not understand natural-language intent itself. Semantic inducement, misleading tool descriptions, and execution drift still require runtime defenses such as StruQ, ACE, and Task Shield \cite{S2_r16,SoKA2APayments}.

\subsection{Information Interaction and Knowledge-Source Trust}
\label{subsec:part5-3-information-interaction-and-knowledge-source-trust}

At the information-interaction stage, the objects of trust are information artifacts such as messages, tool descriptions, retrieval results, memories, execution proofs, and audit materials. Information trustworthiness in open networks concerns not only whether transmission is secure, but also whether provenance, versions, and modification histories can be reconstructed after information is copied, summarized, forwarded, and reused across domains.

\subsubsection{Existing Non-Blockchain Methods}
\label{subsubsec:part5-3-1-existing-non-blockchain-methods}

In communication, protocol, and knowledge-source settings, non-chain methods are strongest at identifying concrete attack surfaces. Agent-in-the-Middle attacks~\cite{s3b_r11} show that an adversary can compromise a multi-agent system by manipulating inter-agent communication without compromising any individual agent. MCP Security Bench~\cite{s3c_r7} reveals protocol attack surfaces such as name collisions, preference manipulation, tool-description injection, and retrieval injection during tool discovery, tool invocation, and response handling. These studies help test the security of concrete protocols and implementations.

The knowledge-source side follows a slightly different logic. RobustRAG~\cite{RobustRAG} represents a path for improving retrieval robustness through isolation and aggregation. Traceback methods such as RAGForensics~\cite{RAGForensics} can help locate poisoned texts and analyze poisoning sources in RAG systems. These methods have direct value for retrieval robustness and pollution localization, and can reduce the probability that polluted content affects concrete tasks.

\subsubsection{Residual Trust Gap}
\label{subsubsec:part5-3-2-residual-trust-gap}

Information artifacts in open agent networks often travel across platforms and are reused multiple times. A tool description may be included in a directory; a retrieved passage may be summarized and written into memory; an execution proof may later serve as evidence for payment or dispute handling. Non-chain methods can detect pollution, locate anomalies, or improve robustness within a single repository or system. However, after information has been copied, compressed, paraphrased, or embedded into other workflows, a single platform cannot easily reconstruct complete provenance on its own. Ordinary logs are also usually stored by one service provider and are not naturally treated as jointly trusted evidence in cross-organizational disputes.

\subsubsection{Blockchain-Based Mechanism Supplement}
\label{subsubsec:part5-3-3-blockchain-based-mechanism-supplement}

Once information flows across platforms, blockchain/DLT is better suited as a shared index for attestation, version tracking, and ex post verification. BlockA2A~\cite{BlockA2A} identifies insecure communication channels as a key vulnerability in A2A interoperability and provides tamper-resistant auditability through a blockchain-anchored ledger. DMAS~\cite{DMAS} supports verifiable interaction cycles, communication integrity, authenticity, non-repudiation, and conditional confidentiality through cryptographic primitives and on-chain operations. MOD-X~\cite{MODX} uses a universal message bus, translation mechanisms, state management, and blockchain security mechanisms so that agent messages across different frameworks, semantics, and protocols can be translated, state-preserved, and verified.

This attestation mechanism does not imply placing all content on-chain. A more reasonable design stores original messages, contexts, retrieved content, and privacy-sensitive materials off-chain, while recording message digests, sender signatures, receiver receipts, execution states, timestamps, and key evidence hashes on-chain \cite{BlockA2A,DMAS}. In a dispute, participants can use on-chain indexes to return to off-chain materials and prove when a material appeared, whether it was modified, who sent it, who confirmed it, and how it entered later execution or settlement.

The objects of on-chain evidence should not be limited to single messages. BetaWeb~\cite{BetaWeb} discusses trustworthy information exchange from the perspective of Agentic Web data governance, while Trust Fabric~\cite{TrustFabric} provides trust mechanisms for cross-domain interoperability through decentralized registries and semantic Agent Cards. Accordingly, tool-description hashes, knowledge-source version digests, directory-update records, and execution-proof indexes can all become objects of on-chain attestation.

This combination of on-chain indexes and off-chain content also addresses the tension between privacy and verifiability. Only digests and states are stored on-chain, avoiding exposure of complete content. Off-chain materials are bound to on-chain records through hashes and signatures, preserving the ability to verify evidence later. Phiri et al.'s axiomatic framework of auditability~\cite{CreatingAuditable} emphasizes integrity, coverage, temporal consistency, and verifiability, and can serve as a conceptual reference for designing on-chain/off-chain evidence. Blockchain supplements a consistent cross-organizational evidence index here, not content security itself.

Thus, on-chain attestation strengthens the verifiability of provenance, versions, and modification histories, rather than the truthfulness or semantic safety of the content itself. RAG robustness, semantic-inducement detection, and knowledge-truth assessment remain the responsibilities of non-chain content-security mechanisms.

\subsection{Collaborative Decision-Making and Group-Robustness Trust}
\label{subsec:part5-4-collaborative-decision-making-and-group-robustness-trust}

At the collaborative decision-making stage, the object of trust is the process through which multiple agents jointly form a judgment, including participants, communication topology, collaboration records, contribution paths, and the basis for dispute handling. Collaboration trustworthiness in open networks cannot be reduced to majority agreement or task-internal consensus, because agreement may arise from correlated errors, malicious influence, or vulnerable topology.

\subsubsection{Existing Non-Blockchain Methods}
\label{subsubsec:part5-4-1-existing-non-blockchain-methods}

Task-internal collaboration robustness already has a set of non-chain tools. IBGP~\cite{s3b_r18} adapts the traditional Byzantine problem into a local or partial consensus problem more suitable for communication-based multi-agent systems. G-Safeguard~\cite{s3b_r16} detects anomalous influence from discourse interaction graphs and conducts topology intervention. Agents Under Siege~\cite{s3b_r15} shows that attackers can optimize prompt propagation paths over specific communication topologies, making single-point guardrails insufficient for covering collaborative propagation processes. MultiAgentBench~\cite{s3b_r14} further indicates that multi-agent collaboration and competition outcomes are affected by task structure and communication modes.

In controlled systems or concrete tasks, these methods can identify malicious participation, topological anomalies, and collaboration failures. They emphasize a common point: group judgment is not simply the sum of individual agent outputs, but depends on who participates, how communication occurs, how information flows, and whether different models and roles provide sufficient diversity.

\subsubsection{Residual Trust Gap}
\label{subsubsec:part5-4-2-residual-trust-gap}

Collaboration in open networks is often cross-organizational, cross-platform, and economically consequential. Task-internal topology detection can identify anomalous influence in a particular collaboration, but it does not easily allow other platforms to inherit the same collaboration history. Local malicious-agent identification can improve the quality of the current system, but it may not produce long-term reputation effects. Majority agreement or on-chain consensus also does not directly represent semantic correctness. Collaboration trustworthiness requires more than whether an answer can be formed at the moment. It also requires records of how a task was delegated, which agents participated, who provided key evidence, where disputes arose, how failure should be compensated, and whether those records can affect future selection.

\subsubsection{Blockchain-Based Mechanism Supplement}
\label{subsubsec:part5-4-3-blockchain-based-mechanism-supplement}

Introducing blockchain/DLT into collaboration first serves to preserve collaboration-process records and to provide a basis for incentive constraints and dispute handling. BlockA2A~\cite{BlockA2A} incorporates Byzantine agents and adversarial prompts into its analysis of A2A interoperability risks and provides process auditability through an on-chain anchored ledger. DMAS~\cite{DMAS} enables multi-party collaboration records to be reviewed after task completion through verifiable interaction cycles and non-repudiation. These records do not guarantee that collaboration results are correct, but they can show how tasks were delegated, which agents participated, and when key interactions occurred, thereby supporting reputation updates, dispute adjudication, and responsibility allocation.

Records alone are not enough. Collaboration governance in open networks also depends on incentive constraints. Agent Exchange~\cite{AgentExchange} uses auction platforms, Agent Hubs, dynamic capability representation, performance tracking, and incentive-compatible value attribution to constrain service selection, team formation, and contribution allocation through market mechanisms. Hu and Rong~\cite{InterAgentTrustModel} incorporate primitives such as brief, claim, proof, stake, reputation, and constraint into a unified framework for expressing capability claims, staking commitments, reputation effects, and collaboration constraints. These mechanisms turn collaboration from low-cost trial-and-error into a process with commitments, records, and consequences.

When collaboration fails or breach occurs, on-chain evidence can enter dispute and compensation mechanisms. Insured Agents~\cite{InsuredAgents} introduces risk pricing, conditional audit access, and decentralized verification, allowing transaction failure or breach to enter insurance compensation and responsibility tracing. In this structure, process records provide the factual basis, staking and reputation increase the cost of breach, and insurance and dispute handling provide mechanisms for loss sharing after failure. Blockchain supplements collaboration not by determining the correctness of conclusions, but by making collaboration processes visible, untrusted behavior costly, and dispute handling evidence-based.

On-chain explorations in task-internal MAS can serve as local mechanism references. BlockAgents~\cite{BlockAgents}, DecentLLMs~\cite{DecentLLMs}, DAO-Agent~\cite{DAOAgent}, and WBFT~\cite{WBFT} respectively address Byzantine attacks, leader single-point fragility, zero-knowledge contribution verification, and weighted BFT consensus in LLM multi-agent coordination. They can improve the transparency and malicious-resilience of specific multi-agent coordination processes, but they are not equivalent to cross-organizational discovery, authorization, transaction, and dispute governance in open A2A/IoA settings.

The easiest misunderstanding in this stage is to treat on-chain records, staking, or insurance as sources of result correctness. They can increase the cost of breach and preserve process evidence, but they cannot turn majority agreement into truth, nor can they eliminate model bias, group hallucination, or correlated errors. Immediate robustness detection still requires non-chain methods such as IBGP and G-Safeguard.

\subsection{Value Settlement and Accountability-Governance Trust}
\label{subsec:part5-5-value-settlement-and-accountability-governance-trust}

At the value-settlement stage, the object of trust is the post-execution record and its consequences, including contribution evidence, payment conditions, revenue allocation, resource consumption, audit trails, reputation updates, and responsibility determination. Settlement trustworthiness in open networks requires different organizations to use the same body of evidence to confirm who did what, what compensation is due, where failure occurred, and how accountability should be assigned.

\subsubsection{Existing Non-Blockchain Methods}
\label{subsubsec:part5-5-1-existing-non-blockchain-methods}

Non-blockchain value-governance methods can evaluate contributions, shape cooperation incentives, and generate audit records. Reputation-Filtered Reward Reshaping~\cite{ReputationFiltered} reduces the impact of low-quality contributions through reputation filtering and reward reshaping. LLM multi-agent credit-assignment research~\cite{LLMCreditAssignment} attempts to decompose team outcomes into individual contributions and provide a basis for reward allocation. Auditability and observability research both emphasize recording and monitoring agent behavior \cite{CreatingAuditable,AgentOps}. Among them, Phiri et al.'s work on auditability~\cite{CreatingAuditable} further requires such records to be recoverable, inspectable, and analyzable.

These studies show that value governance is not merely a matter of payment. It also concerns how contribution is defined, how processes are recorded, and how responsibility is explained. In closed systems or platform-internal collaboration, credit assignment, reputation rewards, and observability logs provide important foundations for evaluation and governance.

\subsubsection{Residual Trust Gap}
\label{subsubsec:part5-5-2-residual-trust-gap}

Settlement evidence in cross-organizational agent networks cannot rely only on records stored by a single platform. If execution records are incomplete, evaluation models can be manipulated, logs can be modified after the fact, or contribution calculation is performed only inside one service provider, then payment, compensation, and responsibility analysis are difficult for other participants to jointly accept. Non-chain methods can help evaluate contributions and generate records, but they cannot by themselves guarantee non-repudiable contribution, executable revenue allocation, provable responsibility chains, or tamper-resistant audit logs. As agents purchase data, pay for computation, sell services, subcontract tasks, and influence reputation, value settlement needs to enter a programmable, contestable, and traceable shared-ledger structure.

\subsubsection{Blockchain-Based Mechanism Supplement}
\label{subsubsec:part5-5-3-blockchain-based-mechanism-supplement}

Value settlement more directly touches the programmable execution capability of blockchain/DLT: payment conditions, task proofs, revenue allocation, and dispute handling can all enter the same state machine. Research on the Agent Economy~\cite{TheAgentEconomy} argues that blockchain provides autonomous AI agents with permissionless participation, trustless settlement, and machine-to-machine micropayments, enabling agents to form relatively independent identity, asset, and settlement foundations. A2A+x402~\cite{A2Ax402} combines on-chain AgentCards with x402 micropayments, linking service discovery, identity authentication, and automatic compensation in cross-organizational invocations. The SoK on A2A payments~\cite{SoKA2APayments} systematizes the four-stage lifecycle of discovery, authorization, execution, and accounting, as well as protocol-level challenges such as weak intent binding and misuse under valid authorization. The SoK on Agentic Commerce security~\cite{SoKAgenticCommerce} builds a security threat framework from dimensions such as market manipulation and compliance.

In this structure, payment is no longer an isolated action after task completion. It is jointly defined by authorization conditions and execution evidence. A directory discovers the service provider, policy mechanisms lock budgets and invocation scopes, execution generates task proofs or acceptance materials, verification triggers payment, and failure enters dispute handling, compensation, and reputation updates. Because these steps may belong to different organizations, on-chain contracts and ledgers provide jointly recognized settlement rules and state records, making settlement less dependent on a single platform's internal database or manual reconciliation.

Concrete settlement mechanisms can be viewed from three directions. TessPay~\cite{TessPay} centers on verify-then-pay and binds payment conditions to verifiable execution results. CPMM~\cite{CPMM} combines x402/H402 micropayments, secure capability discovery, and multi-step capability negotiation, providing a mechanism framework for agent capability pricing and micromarket transactions. Agent TCP/IP~\cite{AgentTCPIP} emphasizes that an agent economy needs binding protocols for inter-agent transactions, allowing services, data, and other value objects to enter executable exchange relationships. Together, these mechanisms show that on-chain settlement is not the automation of traditional payment, but the incorporation of service relationships, authorization scopes, execution proofs, capability pricing, and value objects into programmable rules.

Accountability governance also requires handling failure, breach, and dispute after payment. Insured Agents~\cite{InsuredAgents} introduces risk sharing and conditional audit access in cases of transaction failure or breach, linking loss sharing and dispute handling to execution evidence. Phiri et al.'s work on auditable agentic AI~\cite{CreatingAuditable} provides a conceptual reference for the integrity, coverage, temporal consistency, and verifiability that on-chain/off-chain audit records should satisfy. On-chain mechanisms therefore connect settlement, auditing, compensation, and reputation updates, making who completed what, when it was completed, whether conditions were satisfied, and how failure should be handled into cross-organizational facts that can be verified.

Even if payment and audit evidence enter the chain, service quality itself is not automatically guaranteed, and complex legal responsibility cannot be directly adjudicated by contracts. These issues still require off-chain governance mechanisms such as task acceptance, service-level monitoring, dispute arbitration, and regulatory or judicial processes.

\subsection{Synthesis}
\label{subsec:part5-synthesis}

The five trust crises in open LLM agent networks arise from a change in how trust is organized under open network structures, and they cannot be reduced to an accumulation of isolated security vulnerabilities. Account, permission, logging, evaluation, and payment mechanisms inside a single platform become difficult to use as a common trust foundation once they enter cross-organizational, cross-protocol, dynamically accessible, and transactable agent networks. Against this background, the value of blockchain mainly lies in cross-subject trust coordination: it transforms local security results, interaction records, economic commitments, and accountability evidence into states that multiple parties can jointly read, review, and use for dispute handling.

From this perspective, the role of blockchain/DLT differs across levels. Entity admission and value settlement are closest to on-chain public infrastructure. The former requires identity, capability, endpoint, and reputation to be sedimented as subject states that can be continuously reviewed, while the latter requires task proofs, payment conditions, compensation, and audit evidence to be connected into a cross-organizational settlement loop. Authorization execution and information interaction depend more on a combination of on-chain state/evidence and off-chain runtime defenses: on-chain mechanisms make authorization states, message provenance, version records, and evidence indexes jointly recognizable, while off-chain mechanisms continue to handle natural-language intent, tool semantics, RAG contamination, prompt injection, and execution drift. Collaborative decision-making requires the greatest caution about boundaries. On-chain consensus and staking mechanisms can increase the cost of untrusted behavior and preserve process evidence, but they cannot directly guarantee reasoning correctness or the truth of majority opinions.

The five crises also form a progressive and feedback relationship. Entity admission provides identity and capability evidence, but it cannot guarantee that authorization will not be misused. Authorization policies define action boundaries, but they cannot prove that external information has not been polluted. Interaction evidence supports provenance and disputes, but it does not automatically guarantee multi-party collaboration quality. Collaboration constraints reduce the gains of self-interested or malicious behavior, but settlement, auditing, and accountability mechanisms are still needed for consequences to be recorded. Value settlement then feeds back into reputation, admission, and future market selection. Therefore, the reasonable path for blockchain to empower open agent networks is to build a cross-stage composable trust architecture, rather than stacking isolated on-chain modules at each risk point.

Based on the above analysis, this paper limits the role of blockchain/DLT to reviewable state, non-repudiable evidence, incentive constraints, and programmable settlement. Non-chain methods remain central to runtime security, semantic defense, content robustness, capability evaluation, and local detection; on-chain mechanisms preserve reviewable states and constrain cross-subject consequences. Existing literature remains dominated by architectural proposals, early protocol designs, and preprints. Whether it can truly form trust infrastructure for open agent networks still depends on further validation in deployable systems, unified evaluation, privacy protection, cross-protocol revocation, semantic-attack defense, and regulatory compliance.

\section{Representative Application Scenarios and Deployment Implications}
\label{sec:application_scenarios}

The preceding sections have developed a trust-crisis taxonomy and analyzed blockchain-enabled mechanisms for open agent networks. This section further examines how these trust requirements appear in representative deployment scenarios. The scenarios are not selected as ordinary industry verticals. Rather, they correspond to three deployment boundaries in which agent-network trust becomes operationally consequential: open market exchange, cross-organizational institutional workflows, and cyber-physical collaboration. These boundaries differ in participant openness, authority structure, information lifecycle, decision latency, and accountability requirements. They therefore provide a scenario-level synthesis of the preceding taxonomy rather than a separate list of blockchain-agent applications.

At the scenario level, the relevant question is not whether blockchain can be inserted into an agent system, but which trust states must be recognized, inherited, contested, or settled across multiple parties. Agent service markets and the agent economy represent open exchange among heterogeneous and economically motivated agents, where the central problem lies in the feedback between admission and settlement. Cross-organizational workflow automation represents institutional collaboration across partially trusted organizations, where the core issue is the continuity of authority and provenance across organizational boundaries. Embodied, IoT, and vehicular agent networks represent cyber-physical collaboration under low-latency and safety-critical constraints, where blockchain has a more constrained but analytically clearer role in device-trust anchoring, auditability, incentives, and ex post accountability. This comparison clarifies where blockchain contributes shared state, tamper-evident evidence, programmable constraints, and settlement infrastructure, and where off-chain mechanisms remain necessary for semantic judgment, capability evaluation, runtime monitoring, privacy protection, and safety control.

\subsection{Agent Service Markets and the Agent Economy}
\label{subsec:agent_service_markets}

Agent service markets are representative of the transition from application-internal agents to open economic agents. In this scenario, agents are not only task executors, but also service providers, requesters, evaluators, brokers, and economic actors. They may advertise capabilities, bid for tasks, invoke paid tools, purchase data, subcontract subtasks, receive compensation, and accumulate reputation. Recent proposals on agent exchanges, agentic commerce, and verify-then-pay infrastructures suggest that future agent ecosystems may develop market structures for discovery, pricing, task allocation, evaluation, micropayment, and dispute handling.

The network form in this scenario is an open service-exchange network. A requester agent may discover candidate agents through registries or marketplaces, select them according to capability descriptions and reputation records, and coordinate execution through contractual or protocol-level commitments. Unlike closed multi-agent systems, participants are not predefined by one developer and may not share the same identity provider, evaluation standard, reputation system, or settlement infrastructure. Therefore, the most critical risks arise at the two ends of the task lifecycle: before execution, when the market decides which agents are admissible and credible; and after execution, when the market determines whether the result, contribution, payment, and responsibility records are acceptable.

This scenario mainly instantiates Crisis~1, entity admission and capability trust, and Crisis~5, value settlement and accountability-governance trust. At the admission side, the market must determine whether a discovered agent is a legitimate, unique, and qualified participant, and whether its claimed capability, endpoint, price, and historical performance are credible. At the settlement side, the market must determine whether a task was completed, which agents contributed to the result, whether compensation should be released, and how disputes or failures should affect future reputation. The key amplification mechanism is the feedback loop between admission and settlement: an admission error may distort ranking, routing, and demand signals, while a settlement error may corrupt reputation and influence future admission decisions.

Blockchain is useful in this scenario when market states must be jointly recognized by mutually distrustful participants. The key verifiable states include decentralized identities, credential status, service registration, capability-evidence commitments, task commitments, escrowed payment states, signed evaluation records, reputation updates, and dispute evidence. These states allow the market to move from platform-local records to shared and contestable records. On-chain mechanisms can make identity bindings, endpoint commitments, and credential states independently checkable, make task commitments and payment conditions enforceable once predefined conditions are satisfied, and make settlement outcomes reusable as evidence for later audit and reputation update. The deployment implication is that an agent market should not treat blockchain as a capability oracle, but as a market trust substrate that connects admission, commitment, settlement, and reputation into the same evidentiary loop. Capability benchmarking, model fingerprinting, sandbox evaluation, task-quality assessment, and legal-responsibility adjudication remain off-chain complements.

\subsection{Cross-Organizational Workflow Automation}
\label{subsec:cross_organizational_workflows}

The second scenario represents institutional workflows across supply chains, government-enterprise processes, and cross-departmental data processing. In these settings, agents act on behalf of organizations, departments, regulators, suppliers, logistics providers, auditors, or data custodians. They may verify compliance documents, reconcile invoices, query enterprise databases, coordinate logistics, process regulatory materials, and generate audit reports. This scenario is representative not because it is more open than all other settings, but because it exposes a different deployment constraint: participants are often known and permissioned, yet their internal systems, logs, and policy interpretations are not fully trusted by others.

The network form is a permissioned but multi-organizational agent network. Each organization may maintain its own identity system, access-control policy, data repository, logging mechanism, and compliance procedure. An agent-mediated workflow may involve a chain of delegated actions: a user authorizes an internal agent, the internal agent requests information from another department, a supplier-side agent returns documents, and an auditor agent verifies evidence against policy constraints. At the same time, the output of one agent may become the input, evidence, or compliance basis for another. The central deployment problem is therefore not open admission, but the continuity of authority and provenance across institutional boundaries.

This scenario mainly instantiates Crisis~2, authorization, delegation, and execution-boundary trust, and Crisis~3, information interaction and knowledge-source trust. Authorization risk is amplified because a locally valid permission may become globally inappropriate after multi-hop delegation, cross-domain reinterpretation, or task-scope drift. Information risk is amplified because documents, tool outputs, database records, and intermediate summaries may be copied, transformed, summarized, and reused across institutional boundaries. In such workflows, it is insufficient to know that an action occurred or that a document exists. The network must reconstruct who authorized the action, under what scope, which information was used, which version was relied upon, and how it entered the final decision.

Blockchain can serve as a shared evidence substrate for authorization and provenance in this scenario. The main evidentiary objects include delegation states, authorization scopes, revocation events, data-access receipts, document-version commitments, metadata digests, tool-output attestations, and audit indexes. In a consortium or permissioned ledger, these objects can be jointly maintained as commitments, digests, or indexes, while sensitive business or personal data remain off-chain under access-controlled disclosure. The ledger can provide an independently checkable reference for whether a request is associated with a legitimate delegation chain and whether an information artifact is linked to a verifiable provenance trail. The deployment implication is that blockchain is most valuable here as a cross-organizational evidence layer: it reduces disputes over authorization state, provenance, and auditability, while semantic compliance, policy interpretation, data-quality assessment, privacy review, and prompt-injection defense remain off-chain institutional and technical responsibilities. In this sense, cross-organizational workflows are not primarily about openness, but about preserving authority and evidence continuity when agents act across institutional seams.

\subsection{Embodied, IoT, and Vehicular Agent Networks}
\label{subsec:embodied_iot_vehicular_agents}

The third scenario represents cyber-physical collaboration among embodied agents, IoT devices, connected vehicles, roadside units, drones, robots, edge servers, and intelligent infrastructure. Unlike digital service markets or document-centric workflows, these systems exchange sensor observations, local world models, safety warnings, control suggestions, and collaborative perception results. Vehicle-to-everything cooperative perception is a typical example: vehicles and infrastructure agents may improve perception by sharing observations from different viewpoints, while also facing sensor heterogeneity, pose errors, communication constraints, temporal asynchrony, and safety-critical latency.

The network form is a cyber-physical multi-agent network. Agents observe the physical environment locally and transform distributed observations into collective state estimates or decisions. The deployment boundary is defined by latency, bandwidth, physical uncertainty, and safety control. A vehicle, robot, or industrial controller cannot wait for slow global consensus before braking, avoiding obstacles, or responding to a hazard. However, the same system still requires trustworthy records after an incident, because an unsafe decision may be produced by the interaction of multiple sensors, agents, models, communication links, and infrastructure operators.

The dominant trust concern is the reliability and accountability of the collective decision process, corresponding primarily to Crisis~4, collaborative decision-making and group-robustness trust. This crisis is amplified in cyber-physical networks because group judgment is constrained by communication topology, timing, and physical uncertainty. Malicious nodes, faulty sensors, correlated perception errors, delayed messages, or topology bottlenecks may distort the collective result even when many individual messages appear valid. Studies on multi-agent collaboration and topology-guided security show that interaction structure can shape both performance and attack propagation. In embodied and vehicular settings, such propagation is more consequential because it may affect safety-relevant decisions.

Blockchain has a more constrained but conceptually clearer role in this scenario. It should not be positioned as an online decision mechanism or real-time consensus layer. Instead, it is better understood as an ex post accountability, device-trust, incentive, and audit substrate. The relevant verifiable states include device identities, software or model-version commitments, sensor-data digests, signed collaboration records, event summaries, contribution evidence, reputation updates, and incident-investigation materials. These records can help support post-event accountability, maintenance, insurance, and regulatory review, and may also contribute to incentives for honest sensing and cooperative data sharing when coupled with off-chain validation. The deployment implication is that blockchain should remain outside the hard real-time control loop. Real-time perception, sensor fusion, anomaly detection, secure communication, trusted hardware, and fail-safe control must remain off-chain. Blockchain can strengthen evidence-based accountability and coordination, but it does not replace safety-critical decision logic.

\section{Future Work}

Blockchain-based trust infrastructure for open agent networks remains at an early stage. Existing studies have demonstrated the potential of mechanisms such as identity registration, programmable authorization, on-chain attestation, reputation, staking, insurance, and automatic settlement. However, most work still focuses on local protocols, architectural proposals, or single-point prototypes. Future research should move toward long-term scientific questions: how dynamic trust states should be managed, how agent capabilities can be verified, how autonomous economic behavior should be constrained, how collaboration quality should be evaluated, and how on-chain evidence can support verifiable execution.

This section therefore organizes future work into six interrelated research directions. They still correspond to the five trust crises discussed above, but the emphasis shifts toward problem definition, model abstraction, evaluation methodology, and governance interfaces, rather than conventional performance optimization or system scaling.

\subsection{Dynamic Trust Management and Capability State}

Trust state in open agent networks should not be treated as a one-time registration result. It continuously changes during discovery, invocation, collaboration, settlement, and dispute handling. Future research needs to establish dynamic trust-management models so that identity, service endpoints, capability claims, authorization status, reputation, and revocation information can be updated throughout the agent lifecycle. Ecosystems such as A2A, MCP, ACP, ANP, AgentCard, ANS, ERC-8004, x402, and AP2 are developing in parallel. How identity and service states migrate, synchronize, and expire across protocols will directly affect discovery, delegation, and accountability in open networks.

The core question in this direction is how to extend on-chain state from a static index into an evolvable trust-state machine. DIDs, wallet addresses, AgentCards, service domains, public keys, and organizational identities need clear binding relationships so that network participants know not only which key an agent controls, but also how to distinguish the key controller, service operator, model provider, tool provider, and responsible principal. Revocation mechanisms should also extend from certificates to service endpoints, capability credentials, reputation status, payment eligibility, and data access permissions.

Capability state should also be incorporated into dynamic trust management. A claim that an agent can invoke a class of tools, complete a type of task, or access certain data does not imply that the agent still has that capability under the current model version, permission configuration, data environment, and tool environment. Future research needs to bind capability evaluation results, credential digests, model versions, tool permissions, and revocation states, while distinguishing self-declared capabilities, third-party-certified capabilities, task-observed capabilities, and long-term reputation-based capabilities.

The admission layer must also address long-term issues caused by new and malicious participants. New agents lack historical performance records, and on-chain registration alone is insufficient to establish capability reliability. Once reputation systems become open, they may also be manipulated through rating inflation, Sybil identities, and fabricated interactions. Identity uniqueness constraints, capability evaluation, on-chain reputation, staking, and third-party certification need to be combined to support continuously updated admission mechanisms.

\subsection{Intent Constraints and Verifiable Execution Boundaries}

The long-term problem for the authorization-execution layer is how to keep agent actions continuously constrained by user intent and authorization boundaries. Authorization in open agent networks often originates from natural-language goals, whereas on-chain contracts are better at expressing budget limits, time windows, invocation scopes, payment conditions, and revocation rules. Future research needs to examine how user intent can be transformed into structured authorization objects, and how tool invocation, subdelegation, data access, and payment behavior can remain constrained by such objects.

Verifiable execution boundaries need to connect on-chain authorization states with off-chain runtime checks. The chain can record authorization-object digests, budget states, subdelegation relationships, revocation events, and payment conditions. The off-chain execution environment then needs to determine before tool invocation, subdelegation, and payment whether an action still aligns with the original task objective. This is necessary for combining non-chain intent understanding and task-consistency checking with verifiable on-chain authorization states.

Because an agent's action boundary may change during execution, authorization mechanisms also need to cover dynamic behavior. An agent may discover new services, negotiate new prices, invoke new tools, or adjust its task path during execution. A system cannot require human confirmation at every step, but it also cannot allow unrestricted privilege expansion. Progressive authorization, conditional subdelegation, behavior-triggered revocation, and dynamic budget tightening will therefore become important problems for the authorization-execution layer.

\subsection{Verifiable Execution and Semantic Evidence}

Verifiable execution requires on-chain evidence to support off-chain semantic verification and to go beyond proof of digest existence. Blockchain can prove that a hash, signature, timestamp, or message digest existed at a particular time and has not been tampered with, but it cannot directly prove that a message is true, a tool description is harmless, a knowledge source is reliable, or a RAG result is uncontaminated. Evidence models therefore need to express provenance, version, context, uncertainty, and execution conditions beyond integrity records alone.

This direction needs to distinguish different evidence types and their verification boundaries. Message interaction requires proof of sender, receiver, time, and integrity. Knowledge sources require records of data source, version, and update history. Tool invocation requires input, output, permission, and execution context. Compliance evidence needs to explain whether access, processing, and sharing satisfy specified rules. How message proof, provenance proof, version proof, execution proof, and compliance proof can be represented uniformly, reused across systems, automatically verified in some cases, and reviewed off-chain in others remains a problem for future research.

Finer-grained evidence also creates greater privacy pressure. More detailed records help accountability and dispute handling, but they may expose user intent, task workflows, data content, and interaction relationships. Zero-knowledge proofs, TEEs, selective disclosure, off-chain encrypted storage, and verifiable logs can serve as candidate tools for building evidence structures that disclose minimally while remaining sufficient for audit.

\subsection{Trustworthy Collaboration Evaluation}

Trustworthy collaboration evaluation needs to move from ``recording who participated'' toward ``explaining why a collaborative process produced a particular result.'' Tasks in open agent networks are often completed by multiple agents through dynamic discovery, temporary teaming, subcontracting, voting, debate, negotiation, and multi-hop delegation. On-chain consensus can indicate which agents participated, when they submitted results, and whether some commitment was formed, but it cannot prove that majority opinion is factual or automatically remove correlated hallucination and group-level error.

Beyond outcome trustworthiness, the collaboration layer must handle compositional responsibility. A failed result may arise from upstream data contamination, exaggerated capability claims by an agent, incorrect task decomposition by the delegator, tool invocation failure by the executor, erroneous acceptance by the evaluator, or correlated errors caused by multiple agents using similar models and data sources. Roles, contributions, dependencies, and dispute points in the collaborative process need to be recorded so that responsibility allocation has a reviewable evidentiary basis.

Future research needs to establish evaluation metrics for collaboration quality beyond process records alone. Possible metrics include contribution attributability, diversity of information sources, model correlation, abnormal influence, stability of group decisions, and dispute reviewability. Reputation, staking, voting, and insurance can increase the cost of misbehavior, but governance itself may fail if reputation is inflated, voting is attacked by Sybil identities, or staking costs remain lower than the payoff from manipulation. Manipulation-resistant reputation updates, contribution attribution, staking penalties, and insurance pricing need to be designed together with collusion detection, market-manipulation defense, and governance for resource concentration.

\subsection{Agent Economy and Accountability Loops}

The long-term problems of the Agent Economy go beyond automatic payment. They also include how to constrain autonomous economic behavior, how to price capabilities, how to handle breach, and how to feed economic consequences back into reputation and admission. On-chain payment can prove that funds have been transferred, but it cannot automatically prove that a service has been completed, that quality requirements have been met, or that user intent has been satisfied. Mechanisms such as verify-then-pay, task proofs, escrowed payment, insurance compensation, and dispute arbitration should be organized into accountability-governance chains rather than treated as single payment actions.

To support such binding, the settlement layer needs to distinguish different proof objects. Task proof concerns whether a task was executed. Outcome proof concerns whether the result satisfies agreed conditions. Settlement proof concerns whether payment and resource states are consistent. Liability proof concerns failure causes, responsible principals, and subsequent handling. Different proof objects may have different degrees of automation: some can be triggered by contracts, while others still need off-chain arbitration, insurance, or legal processes.

Regulatory compliance and legal responsibility cannot be fully delegated to smart contracts. Accountable principals in open agent networks may include users, agent developers, model providers, tool providers, on-chain protocols, data providers, and deployment platforms. On-chain evidence needs interfaces to off-chain governance so that these records can be understood by regulators, arbitration bodies, insurers, and legal processes, rather than remaining internally consistent only within technical systems.

\subsection{Evaluation Benchmarks and Testbeds}

All of the above directions ultimately require unified evaluation and testbeds. Existing experiments often evaluate only throughput, latency, contract cost, cryptographic overhead, or a single attack scenario. However, real risks in open agent networks span identity, capability, authorization, execution, collaboration, and settlement. A more suitable benchmark should cover agent registration and discovery, capability claims and evaluation, user authorization and subdelegation, inter-agent communication and task collaboration, task proof and payment settlement, dispute handling, and reputation update.

Evaluation metrics should also expand from traditional blockchain performance to agent-network trust effects. In addition to TPS, latency, gas cost, and storage overhead, they should include dynamic trust-update quality, identity verifiability, revocation propagation time, capability-verification accuracy, authorization-deviation detection rate, execution-evidence completeness, contestability of task outcomes, accountability-attribution accuracy, privacy leakage, cross-protocol interoperability, and human audit cost. Only when these metrics become systematic can blockchain mechanisms move from conceptual proposals toward comparable, reproducible, and deployable trust infrastructure.

\section{Conclusion}

This survey examined blockchain-empowered trustworthy AI agent networks from network and lifecycle perspectives. We characterized the trust challenges of open agent ecosystems along five dimensions: entity admission and capability, authorization and execution boundaries, information and provenance, collaborative decision making, and settlement and accountability. We also analyze how decentralized identities, verifiable credentials, registries, smart contracts, provenance mechanisms, reputation, staking, and payments can provide shared state, verifiable commitments, and enforceable consequences across organizational boundaries.

Blockchain, however, does not directly guarantee capability quality, semantic truth, safe tool use, task correctness, privacy, or robust collective reasoning. Its value is greatest when mutually distrustful participants require persistent shared records, programmable coordination, or contestable settlement; otherwise, conventional access control, signed logs, trusted execution, runtime monitoring, and centralized infrastructures may be more appropriate. Trustworthy agent networks will therefore require hybrid architectures in which blockchain supports cross-organizational trust and economic coordination, while off-chain mechanisms secure reasoning, execution, privacy, real-time interaction, and physical control. Future research should prioritize dynamic trust management, verifiable capability claims, privacy-preserving evidence, robust collaboration evaluation, manipulation-resistant incentives, interoperability, and lifecycle-oriented benchmarks.

{\footnotesize
	\bibliographystyle{IEEEtran}
	\bibliography{ref.bib}
}

\vfill

\end{document}